\documentclass[a4paper,12pt]{article}
\usepackage{amsmath,amssymb,bm,amsthm}
\usepackage{mathrsfs}
\usepackage[dvipdfmx]{color,graphicx}
\usepackage{enumerate}

\title{On the Uniqueness and Stability of the Equilibrium Price in Quasi-Linear Economies}
\author{Yuhki Hosoya\thanks{1-50 1601 Miyamachi, Fuchu, Tokyo 183-0023, Japan.} {}\thanks{TEL: +81-90-5525-5142, E-mail: ukki(at)gs.econ.keio.ac.jp}\\ Faculty of Economics, Chuo University\\ 742-1 Higashinakano, Hachioji, Tokyo 192-0393, Japan.}
\date{}
\makeatletter
\renewenvironment{proof}[1][\proofname]{\par
  \pushQED{\qed}%
  \normalfont \topsep6\p@\@plus6\p@\relax
  \trivlist
  \item\relax
  {\bfseries
  #1\@addpunct{.}}\hspace\labelsep\ignorespaces
}{%
  \popQED\endtrivlist\@endpefalse
}
\makeatother

\theoremstyle{definition}
\newtheorem{prop}{Proposition}
\newtheorem{thm}{Theorem}
\newtheorem{lem}{Lemma}
\newtheorem{cor}{Corollary}

\begin{document}
\maketitle
\begin{abstract}
In this paper, we show that if every consumer in an economy has a quasi-linear utility function, then the normalized equilibrium price is unique, and is locally stable with respect to the t\^atonnement process. Our study can be seen as that extends the results in partial equilibrium theory to economies with more than two dimensional consumption space. Moreover, we discuss the surplus analysis in such economies.

\vspace{12pt}

\vspace{12pt}

\end{abstract}

\section{Introduction}
How many competitive equilibrium prices are there? Since Arrow and Debreu (1954) showed that there is at least one equilibrium price, this issue has been of great interest to economists. If the equilibrium price is unique, then it would greatly increase the predictive accuracy of a model. However, even in textbook-level examples, the economy can have multiple equilibria (e.g., see exercise 15.B.6 of Mas-Colell et al. (1995)). On this issue, economists had divided into two positions. The first position tries to guarantee the uniqueness of the equilibrium price by making strong assumptions about the economy. The second position gives up the uniqueness of equilibrium prices. In particular, the positions of Arrow and Debreu were clearly divided on this matter.

Gerard Debreu took the latter position. He said that all known conditions for guaranteeing the uniqueness of equilibrium prices are ``exceedingly strong'' (Debreu, 1972). On the other hand, he proved the Sonnenschein--Mantel--Debreu theorem (Debreu, 1974). This theorem implies that any compact set in the positive orthant can be included in the set of equilibrium prices when we prohibit to make any additional assumptions othe than the usual, widely accepted assumptions of the economy. Thus, without any assumptions that he claimed to be ``exceedingly strong'', we know nothing about the number of equilibria. The number of equilibria may be one, a million, or infinity. To solve this problem, Debreu (1970) treated perturbations of the economy by initial endowment vectors, and showed that in ``almost all'' economies, the number of equilibria is at least not infinite. Later, this result has been refined and developed into the theory of regular economy.

In contrast, Kenneth Arrow took the former position. He treated this problem in combination with another traditional problem, called the theory of the t\^atonnement process. In the 19th century, Walras (1874) gave the following explanation for why the equilibrium price is realized in a competitive equilibrium model. First, when the price is higher than the equilibrium price, supply will be high and demand will be low. This situation is called a state of excess supply. In this situation, a lot of goods remain unsold and inventory increases. Therefore, the price will go down. On the other hand, if the price is lower than the equilibrium price, the opposite will occur: demand will be high and supply will be low. This situation is called a state of excess demand. In this instance, there will be many sellouts and the price will be high. As a result, the equilibrium price would attract the actual price, and the economy tends to trade by the equilibrium price. This idea was refined and expressed in differential equations, called the {\bf t\^atonnement process}.

Arrow et al. (1959) showed that if the excess demand function is gross substitute, then any equilibrium price is a globally stable steady state with respect to the t\^atonnement process. Because there can be only one globally stable steady state, they simultaneously showed that there is only one equilibrium price. The next problem is determining the conditions of the economy under which the excess demand function satisfies the gross substitution. However, this problem has not yet been resolved. As a result, Debreu's view that ``there is a little that can be said about the number of equilibria in the general environment'' is now common among economists.

There is another theory of equilibrium besides the general equilibrium theory, namely, the partial equilibrium theory. The partial equilibrium theory has a foundation in general equilibrium theory, where there are only two types of goods, numeraire good and traded good, and consumers' utility functions must be quasi-linear. In the partial equilibrium model, it is easy to show by drawing a diagram that there is only one equilibrium price and that this price is globally stable. In other words, in a quasi-linear economy with two commodities, it is expected that the equilibrium price is unique, and is globally stable with respect to the t\^atonnement process.

The purpose of this paper is to extend this result to a quasi-linear economy with more than two commodities. That is, the aim of this study is to determine whether the above result holds when considering a general equilibrium model in which the utility remains quasi-linear and the dimension of the consumption space may be greater than two. The results are as follows: first, the equilibrium price is unique up to normalization in an economy where all consumers have quasi-linear utility functions. Second, this equilibrium price is locally stable with respect to the t\^atonnement process (Theorem 1). As expected from partial equilibrium theory, if the number of commodities is two, the equilibrium price is globally stable (Proposition 3). However, if the number of commodities is greater than two, then the global stability is not derived in this paper. This is related to the inherent difficulty of quasi-linear economies: see our discussion in subsection 3.1.

We stress that our main result is independent of traditional results concerning the uniqueness of the equilibrium price. In particular, Theorem 1 is independent of results for economies whose excess demand function is gross substitute. It is often misunderstood that, in a quasi-linear economy, although demand functions are always gross substitute, the excess demand function may not be gross substitute.\footnote{See the last two paragraphs in section 2.} Therefore, results using gross substitution cannot be used in our study. Instead, we construct the proof using the index theorem. The index of any equilibrium price in a smooth quasi-linear economy is $+1$, and thus, any such economy is regular. For a non-smooth quasi-linear economy, we use a method of approximation using a mollifier.

This paper also discusses surplus analysis. As in the partial equilibrium theory, the consumer's surplus can be defined for a quasi-linear economy, and can be calculated using only the aggregated demand function. The amount of surplus coincides with the increase in the sum of utilities in the trade of this market. (Theorem 2) This result may be applicable to various applied research.

The structure of this paper is as follows. In section 2, we first define an economy, and then define the type of economy we call a ``quasi-linear economy''. There are two types of quasi-linear economy. One permits negative consumption with respect to the numeraire good,\footnote{This is one of the traditional treatments of a quasi-linear utility function. For example, see the definition of the quasi-linear preference in Mas-Colell et al. (1995).} and the other assumes that the initial endowment of the numeraire good is sufficiently large. Using these preparations, we prove the main result. Section 3 argues several topics concerning our main result. In subsection 3.1, we discuss the difficulty of deriving global stability. In subsection 3.2, we consider the consumer's surplus in this general equilibrium setup. In subsection 3.3, we examine the relationship between the results of this paper and related research. Section 4 includes the conclusion. The proofs are given in section 5.

\section{Results}

\subsection{Preliminary: General Setups of Economies}
In this paper, for $x,y\in \mathbb{R}^K$, $x\ge y$ means that $x_i\ge y_i$ for every $i\in \{1,...,K\}$, and $x\gg y$ means that $x_i>y_i$ for every $i\in \{1,...,K\}$. Define the sets $\mathbb{R}^K_+=\{x\in \mathbb{R}^K|x\ge 0\}$ and $\mathbb{R}^K_{++}=\{x\in \mathbb{R}^K|x\gg 0\}$ as usual. If $K=1$, then we abbreviate this symbol, and simply write these sets as $\mathbb{R}_+$ and $\mathbb{R}_{++}$. The notation $e_j$ denotes the $j$-th unit vector.

Let $L\ge 2$. We call the septuplet 
\[E=(N,M,(\Omega_i)_{i\in N},(U_i)_{i\in N},(\omega^i)_{i\in N},(Y_j)_{j\in M},(\theta_{ij})_{i\in N,j\in M})\]
an {\bf economy} if,
\begin{enumerate}[(1)]
\item $N=\{1,...,n\}$ is a finite set of consumers, and $M=\{1,...,\mu\}$ is a finite set of producers. We admit $\mu=0$ and in this case, this economy is called a {\bf pure exchange economy}.

\item $U_i:\Omega_i\to \mathbb{R}$ denotes the utility function of the $i$-th consumer, where the set $\Omega_i\subset \mathbb{R}^L$ denotes the set of all possible consumption plans for the $i$-th consumer,

\item $\omega^i\in \Omega_i$ denotes the initial endowment of the $i$-th consumer,

\item $Y_j\subset \mathbb{R}^n$ denotes the production set of the $j$-th producer, and

\item $\theta_{ij}\ge 0$ denotes the share of the $i$-th consumer for the $j$-th producer.
\end{enumerate}
We always assume that $\sum_{i\in N}\theta_{ij}=1$ for all $j\in M$.

Consider the following utility maximization problem:
\begin{align}
\max~~~~~&~U_i(x),\nonumber \\
\mbox{subject to. }&~x\in \Omega_i,\label{UMP}\\
&~p\cdot x\le m,\nonumber
\end{align}
where $p\gg 0$ and $m>0$. Let $f^i(p,m)$ denote the set of all solutions to (\ref{UMP}). This set-valued function $f^i(p,m)$ is called the {\bf demand function} of consumer $i$. 

Next, consider the following profit maximization problem:
\begin{align}
\max~~~~~&~p\cdot y,\nonumber\\
\mbox{subject to. }&~y\in Y_j, \label{PMP}
\end{align}
where $p\gg 0$. Let $y^j(p)$ denote the set of all solutions to (\ref{PMP}). This set-valued function $y^j(p)$ is called the {\bf supply function} of producer $j$. Moreover, the value of (\ref{PMP}) is denoted by $\pi^j(p)$, which is called the {\bf profit function}.

We write
\[m^i(p)=p\cdot \omega^i+\sum_{j\in M}\theta_{ij}\pi^j(p).\]
For a given demand function $f^i$ and supply functions $y^1,...,y^m$, define
\[X^i(p)=f^i(p,m^i(p))-\omega^i.\]
This function is called the {\bf excess demand function of consumer $i$}. Define
\[X(p)=\sum_{i\in N}X^i(p),\]
and
\[\zeta(p)=X(p)-\sum_{j\in M}y^j(p).\]
This set-valued function is called the {\bf excess demand function} in this economy. We call $p^*\in \mathbb{R}^L_{++}$ an {\bf equilibrium price} if $0\in \zeta(p^*)$.

Note that, in the usual economy, the excess demand function is homogeneous of degree zero: that is, $\zeta(ap)=\zeta(p)$ for any $a>0$. Therefore, if $p^*$ is an equilibrium price, then $ap^*$ is also an equilibrium price for any $a>0$, which implies that the set of equilibrium prices must not be a singleton. Hence, for arguing the uniqueness of the equilibrium price, normalization must be needed. We say that the equilibrium is {\bf unique up to normalization} if there exists an equilibrium price $p^*$ such that every equilibrium price is proportional to $p^*$.

If the demand function $f^i$ is single-valued and differentiable at $(p,m)$, then we can define
\[s^i_{jk}(p,m)=\frac{\partial f^i_j}{\partial p_k}(p,m)+\frac{\partial f^i_j}{\partial m}(p,m)f^i_k(p,m).\]
The $L\times L$ matrix-valued function $S_{f^i}(p,m)=(s^i_{jk}(p,m))_{j,k=1}^L$ is called the {\bf Slutsky matrix}.

Choose an {\bf adjustment coefficient} $a\in \mathbb{R}^L_{++}$, and consider the following differential inclusion:
\begin{equation}\label{TP}
\dot{p}_{\ell}(t)\in a_{\ell}\zeta_{\ell}(p(t)),\ p_{\ell}(0)=p_{0\ell},\ \ell\in \{1,...,L\},
\end{equation}
where $p_0\in \mathbb{R}^L_{++}$. This inclusion is called the {\bf t\^atonnement process} of the economy $E$ with the adjustment coefficient $a$. We call a set $I\subset \mathbb{R}$ an {\bf interval} if it is convex and contains at least two different points. A function $p:I\to \mathbb{R}^L_{++}$ is called a {\bf solution} to (\ref{TP}) if $I$ is an interval including $0$, $p(t)$ is absolutely continuous on any compact set $C\subset I$, $p(0)=p_0$, $\zeta(p(t))\neq \emptyset$ for all $t\in I$, and
\[\dot{p}_{\ell}(t)\in a_{\ell}\zeta_{\ell}(p(t)),\]
for all $\ell\in \{1,...,L\}$ and almost all $t\in I$. It is well known that if $\zeta$ is single-valued and continuous, then every solution to (\ref{TP}) is continuously differentiable.

An equilibrium price $p^*$ is said to be {\bf locally stable} if there exists an open neighborhood $U$ of $p^*$ such that 1) for every $p_0\in U$, there exists a solution $p:I\to \mathbb{R}^L_{++}$ to (\ref{TP}) such that $\mathbb{R}_+\subset I$, and 2) for every solution $p:I\to \mathbb{R}^L_{++}$ to (\ref{TP}) such that $\mathbb{R}_+\subset I$, $\lim_{t\to \infty}p(t)=\alpha p^*$ for some $a>0$. If we can choose $U=\mathbb{R}^L_{++}$, then $p^*$ is said to be {\bf globally stable}.

Define the set $Y=\sum_{j\in M}Y_j$. This set is called the {\bf aggregate production set}.

Let $(x,y)=(x^1,...,x^n,y^1,...,y^m)\in \mathbb{R}^{(n+\mu)L}$. We call this vecter a {\bf feasible allocation} in $E$ if 1) $x^i\in \Omega_i$ for all $i\in N$, 2) $y^j\in Y_j$ for all $j\in M$, and 3) $\sum_{i\in N}x^i=\sum_{i\in N}\omega^i+\sum_{j\in M}y^j$. Let $\bar{A}_{\omega}$ denote the set of all feasible allocations in the economy $E$. We call this set the {\bf attainable set} for this economy. Later, we write $\bar{A}_{\hat{\omega}}$ to denote the set of all feasible allocations in the economy $E$, where $\omega=(\omega^1,...,\omega^n)$ is replaced with $\hat{\omega}=(\hat{\omega}^1,...,\hat{\omega}^n)$.

Finally, throughout this paper, the symbol $Dh(x)$ denotes the Fr\'echet derivative of $h$ at $x$ and $\nabla h(x)$ denotes the transpose of $Dh(x)$.

\subsection{Economies with Quasi-Linear Environments}

We make certain assumptions regarding economies with quasi-linear environments. However, there are two styles of quasi-linear environments. The first style is that of Mas-Colell et al. (1995), in which the consumption space $\Omega_i$ is assumed to be $\mathbb{R}^{L-1}_+\times \mathbb{R}$. That is, in this style, it is possible to consume the negative amounts of the numeraire good.\footnote{If we consider that $x_L$ denotes the amount of money, negative values of $x_L$ indicate a debt.} In the second style, we assume that $\Omega_i=\mathbb{R}^n_+$ as usual, and the initial endowment of the numeraire good is sufficiently large for every consumer. We treat both styles, and thus we separate the assumptions for these environments.

Throughout this paper, for every $x\in \mathbb{R}^L$, $\tilde{x}$ denotes $(x_1,...,x_{L-1})\in \mathbb{R}^{L-1}$. Here, we make assumptions on economies with quasi-linear environments.

\vspace{12pt}
\noindent
{\bf Assumption F}. For every $i\in N$, $\Omega_i=\mathbb{R}^{L-1}_+\times\mathbb{R}$, and the function $U_i$ can be written as follows:
\begin{equation}
U_i(x)=u_i(\tilde{x})+x_L,\label{QLU}
\end{equation}
where $u_i$ is continuous, nondecreasing, and concave on $\mathbb{R}^{L-1}_+$, and is twice continuously differentiable on $\mathbb{R}^{L-1}_{++}$. Moreover, $\nabla u_i(\tilde{x})\gg 0$, and the Hessian matrix $D^2u_i(\tilde{x})$ is negative definite for every $\tilde{x}\in \mathbb{R}^{L-1}_{++}$.\footnote{This assumption is equivalent to the non-zero Gaussian curvature requirement of Debreu (1972) for every indifference hypersurface of $U_i$ passing through the interior of $\Omega_i$. Debreu (1972) showed that this requirement is equivalent to the differentiability of the demand function at any price and money such that every coordinate of the demand is positive.}

\vspace{12pt}
\noindent
{\bf Assumption S1}. For every $i\in N$, $\Omega_i=\mathbb{R}^L_+$ and the function $U_i$ can be written as (\ref{QLU}), where $u_i$ is continuous, nondecreasing, and concave on $\mathbb{R}^{L-1}_+$, and is twice continuously differentiable on $\mathbb{R}^{L-1}_{++}$. Moreover, $\nabla u_i(\tilde{x})\gg 0$ and the Hessian matrix $D^2u_i(\tilde{x})$ is negative definite for every $\tilde{x}\in \mathbb{R}^{L-1}_{++}$.

\vspace{12pt}
\noindent
{\bf Assumption S2}. For $\hat{\omega}^i=(\omega^i_1,...,\omega^i_{L-1},0)$, define
\begin{equation}\label{ALPHA}
\alpha_i=\sup\{u_i(\tilde{x}^i)-u_i(\tilde{\omega}^i)|(x,y)\in \tilde{A}_{\hat{\omega}},\ U_j(x^j)\ge U_j(\hat{\omega}^j)\mbox{ for all }j\in N\setminus\{i\}\}.
\end{equation}
Then, $\omega^i_L>\alpha_i$ for every $i\in N$.\footnote{The finiteness of $\alpha_i$ is confirmed later. See Proposition 1.}

\vspace{12pt}
\noindent
{\bf Assumption Q}. For every $i\in N$, $\tilde{p}\in \mathbb{R}^{L-1}_{++}$, and $m>0$, the following problem
\begin{align}
\max~~~~~&~u_i(\tilde{x})\nonumber \\
\mbox{subject to. }&~\tilde{x}\in \mathbb{R}^{L-1}_+,\label{QLU2}\\
&~\tilde{p}\cdot \tilde{x}\le m\nonumber 
\end{align}
has an inner solution $\tilde{x}^*\gg 0$. Moreover, $\nabla u_i(\mathbb{R}^{L-1}_{++})=\mathbb{R}^{L-1}_{++}$. In addition, if $u_i(\tilde{x})>u_i(0)$, then $u_i$ is strictly increasing on $\tilde{x}+\mathbb{R}^{L-1}_+$.\footnote{To guarantee the existence of an inner solution, it is usually assumed that the closure of the indifference hypersurface in $\mathbb{R}^{L-1}_{++}$ does not intersect $\mathbb{R}^{L-1}_+\setminus \mathbb{R}^{L-1}_{++}$. Although many typical functions such as $u_i(x)=(x_1x_2)^{1/3}$ satisfy this assumption, a number of functions such as $u_i(x)=\sqrt{x_1}+\sqrt{x_2}$ are excluded. Because we want to deal with these functions simultaneously, we avoid using the above assumption and directly assume the existence of an inner solution. Note that, Assumption Q is satisfied by both functions mentioned above.}

\vspace{12pt}
\noindent
{\bf Assumption U}. $\omega^i\in \mathbb{R}^L_+\setminus \{0\}$ and $\sum_{i\in N}\omega^i\gg 0$.

\vspace{12pt}
\noindent
{\bf Assumption P}. For every $j\in M$, $Y_j\cap \mathbb{R}^L_+=\{0\}$, $Y_j$ is a closed and convex set including $-\mathbb{R}^L_+$, and $Y\cap (-Y)=\{0\}$. Moreover, if $y^1,y^2\in Y_j$, $y^1\neq y^2$, $0<t<1$, and $y^1,y^2\notin -\mathbb{R}^n_+$, then $(1-t)y^1+ty^2$ is in the interior of $Y_j$.\footnote{The first assumption means the possibility of inaction and no-free-lunch condition. The second assumption implies the free disposal property, that is, $Y_j=Y_j-\mathbb{R}^L_+$. See Problem 5.B.5 of Mas-Colell et al. (1995). The third assumption states that the aggregate production set is irreversible. The fourth assumption indicates that this production set is decreasing returns to scale.}

\vspace{12pt}
In this paper, we call an economy $E$ that satisfies Assumptions F, Q, U, and P a {\bf first-type quasi-linear economy}, and that satisfies Assumptions S1, S2, Q, U, and P a {\bf second-type quasi-linear economy}, respectively. We call $E$ a {\bf quasi-linear economy} if it is either a first-type quasi-linear economy or a second-type quasi-linear economy.

The following proposition is needed to justify Assumption S2.

\begin{prop}\label{PROP1}
If Assumptions S1, Q, U, and P hold, then $\alpha_i$ in (\ref{ALPHA}) is finite.
\end{prop}

The following proposition is also needed, but the proof is a little more difficult than the usual case.

\begin{prop}\label{PROP2}
Suppose that $E$ is a quasi-linear economy, and $\zeta$ is the excess demand function of this economy. Then, this function $\zeta$ is single-valued,\footnote{That is, if $\zeta(p)\neq \emptyset$, then it is a singleton.} and satisfies the following {\bf Walras' law}
\begin{equation}\label{WL2}
p\cdot \zeta(p)=0,
\end{equation}
and the {\bf homogeneity of degree zero}
\begin{equation}\label{HDZ2}
\zeta(ap)=\zeta(p)\mbox{ for all }a>0.
\end{equation}
Moreover, there is at least one equilibrium price $p^*$ in this economy.\footnote{Note that, the existence of the equilibrium prices is not trivial because in the first-type quasi-linear economy, the consumption sets are not bounded from below. In particular, this result is not just a corollary of the main theorem of Arrow and Debreu (1954).}
\end{prop}

\subsection{Main Result}
Suppose that $E$ is a quasi-linear economy, and recall the t\^atonnement process
\begin{equation}\label{TP2}
\dot{p}_{\ell}(t)=a_{\ell}\zeta_{\ell}(p(t)),\ p_{\ell}(0)=p_{0\ell},
\end{equation}
where $a\gg 0$ and $p_0\in \mathbb{R}^L_{++}$. Because the excess demand function $\zeta$ is single-valued in quasi-linear economies, the t\^atonnement process is not a differential inclusion, but an ordinary differential equation. Thus, we can use the usual techniques for ordinary differential equations. In particular, any solution $p:I\to \mathbb{R}^L_{++}$ to (\ref{TP2}) is continuously differentiable.

We now complete the preparation of our main result.

\begin{thm}\label{THM1}
Suppose that $E$ is a quasi-linear economy, and $\zeta$ is the excess demand function of this economy. Then, the equilibrium price is unique up to normalization. Moreover, for any adjustment coefficient $a\gg 0$, this equilibrium price is locally stable.
\end{thm}

We make an important note on the independence between this uniqueness and stability result and the traditional one using {\bf gross substitution} of the excess demand function treated in Arrow et al. (1959). Although often misunderstood, the gross substitution of the demand function is a different property from that of the excess demand function. The demand function for each consumer in a quasi-linear economy must be gross substitute, but nontheless, the excess demand function in such an economy is not necessarily gross substitute.

To understand this point, recall the definition of gross substitution. The excess demand function is gross substitute if $\zeta_i$ is increasing in $p_j$ for all $i,j\in \{1,...,L\}$ such that $i\neq j$. Now, consider a first-type quasi-linear pure exchange economy in which $N=\{1\}$ and $L=2$. Let $h(y)=(u_1')^{-1}(y)$. By Lagrange's multiplier rule,
\[\zeta_1(p)=h(p_1/p_2)-\omega^1_1,\ \zeta_2(p)=[p\cdot \omega^1-p_1h(p_1/p_2)]/p_2-\omega^1_2.\]
Therefore,
\[\frac{\partial \zeta_2}{\partial p_1}(p_1,1)=\omega^1_1-p_1h'(p_1)-h(p_1).\]
If $-p_1h'(p_1)>h(p_1)$ for some $p_1$, then $\frac{\partial \zeta_2}{\partial p_1}(p_1,1)<0$ when $\omega^1_1$ is sufficiently small, and thus $\zeta$ is not gross substitute. Note that $h'(y)=(u''(h(y)))^{-1}<0$. Because Assumptions F and Q only state that $u_1'(x)>0$, $u_1''(x)<0$, and $(u_1')(\mathbb{R}_{++})=\mathbb{R}_{++}$, there is no assumption on this economy that prohibits $-yh'(y)>h(y)$. Hence, even in this simple case, the excess demand function may not be gross substitute.

\section{Discussion}

\subsection{Remarks on Global Stability of Equilibria}
We want to show the global stability of an equilibrium price $p^*$. That is, we want to show that {\bf for every} $p_0\in \mathbb{R}^L_{++}$, there exists a solution $p(t)$ to (\ref{TP2}) defined on $\mathbb{R}_+$, and for such a solution, $\lim_{t\to \infty}p(t)$ exists and is proportional to $p^*$. This cannot be proved because we do not assume that $y^j(p)$ is always defined, and thus if $p_0$ is sufficiently far from the equilibrium price, then $\zeta(p_0)$ may be undefined. This problem can be avoided by strengthening Assumption P or, more simply, considering a pure exchange economy. Actually, we can prove the following proposition.

\begin{prop}\label{PROP3}
If $L=2$, then for every quasi-linear economy in which $\zeta(p)$ is defined on $\mathbb{R}^2_{++}$, the equilibrium price is globally stable.
\end{prop}

Unfortunately, this result is verified only when $L=2$. If $L\ge 3$, then there are two hard difficulties that cannot easily be overcome.

First, suppose that $p(t)$ is a solution to (\ref{TP2}) for some $p_0\in \mathbb{R}^L_{++}$ defined on $\mathbb{R}_+$. We want to show that $\lim_{t\to \infty}p(t)$ exists. However, this problem is difficult if $p_0$ is too far from the half-line $\{ap^*|a>0\}$, and we cannot prove this result. For example, we cannot exclude the possibility that the trajectory of $p(t)$ consists of a cycle. If $L=2$, then we can show that such a case vanishes using the intermediate value theorem for $p_1\mapsto \zeta(p_1,1)$. However, if $L\ge 3$, this logic is broken.

The second problem is more serious. That is, we cannot show the existence of a solution $p(t)$ to (\ref{TP2}) defined on $\mathbb{R}_+$. The essential problem is the following: because $\mathbb{R}^L_{++}$ is open, $C\subset \mathbb{R}^L_{++}$ may be not compact even if it is bounded and closed in the relative topology of $\mathbb{R}^L_{++}$. The existence of a solution to (\ref{TP2}) whose trajectory is included in $\mathbb{R}^L_{++}$ is called the `viability problem'. To solve this viability problem, we can use the {\bf strong boundary condition} of the excess demand function. Consider a pure exchange economy $E$ such that the excess demand function $\zeta$ is a single-valued function. We say that $\zeta$ satisfies the strong boundary condition if and only if for every sequence $(p^k)$ in $\mathbb{R}^L_{++}$ such that $p^k\to p\in \mathbb{R}^L_+\setminus (\mathbb{R}^L_{++}\cup \{0\})$ as $k\to \infty$,
\[\sum_{j:p_j=0}\zeta_j(p^k)\to +\infty.\]
Suppose that $\Omega_i$ is bounded from below. Then, by Walras' law, this condition is equivalent to the usual boundary condition, and thus $\zeta$ usually satisfies this condition. Theorem 7 of Hosoya and Yu (2013) states that if $\zeta$ is single-valued, continuous, and homogeneous of degree zero, and it satisfies Walras' law and the strong boundary condition, then there exists a solution to (\ref{TP2}) defined on $\mathbb{R}_+$. Therefore, if the economy $E$ is second-type quasi-linear, then we can solve the viability problem positively.

However, in first-type quasi-linear economies, there is no known method for solving the viability problem. Therefore, it is hard to prove the existence of a solution $p(t)$ to (\ref{TP2}) defined on $\mathbb{R}_+$, and thus the global stability is also difficult to verify.

\subsection{On the Consumer's Surplus}
One virtue of the two-commodity quasi-linear economy is the ability to calculate the change of consumer's utility from the aggregated demand curve. That is, such an economy can be described by a partial equilibrium model, and we can calculate the {\bf consumer's surplus} instead of the utility function directly. It is known that a change in the consumer's surplus coincides with a change in the sum of the utility of consumers in a quasi-linear economy with $L=2$. We can extend this result for the case in which $L\ge 3$.

To simplify the arguments, we assume that $E$ is a first-type quasi-linear economy. We show in the proof section that the domain of $f^i$ can be extended to $\mathbb{R}^L_{++}\times \mathbb{R}$. Thus, we assume here that the domain of $f^i$ is $\mathbb{R}^L_{++}\times \mathbb{R}$. Suppose that $\tilde{p},\tilde{q}\in \mathbb{R}^{L-1}_{++}$. Define the following function:
\[V_i(\tilde{p},\tilde{q})=\int_0^1\tilde{f}^i(c(t),1,m)\cdot c'(t)dt,\]
where $c:[0,1]\to \mathbb{R}^L_{++}$ is a continuously differentiable function such that $c(0)=\tilde{p}$ and $c(1)=\tilde{q}$. The following theorem holds.

\begin{thm}\label{THM2}
$V_i(\tilde{p},\tilde{q})$ is independent of the choice of $c(t)$ and $m$. Moreover, if we set $p_L=q_L=1$, then for any $m$,
\[V_i(\tilde{p},\tilde{q})=u_i(\tilde{f}^i(q,m))-u_i(\tilde{f}^i(p,m)).\]
\end{thm}

Let $D(\tilde{p})$ be the aggregate demand function; that is, for $\ell\in \{1,...,L-1\}$,
\[D_{\ell}(\tilde{p})=\sum_{i\in N}\tilde{f}^i_{\ell}(\tilde{p},1,0).\]
Then, we obtain the following corollary.

\begin{cor}\label{COR1}
Choose any $\tilde{p},\tilde{q}\in \mathbb{R}^{L-1}_{++}$, and suppose that $c:[0,1]\to \mathbb{R}^{L-1}_{++}$ is a continuously differentiable function such that $c(0)=\tilde{p}$ and $c(1)=\tilde{q}$, and $p_L=q_L=1$. Then, for all $m_1,...,m_n$,
\begin{equation}\label{CS}
\int_0^1D(c(t))\cdot c'(t)dt-[\tilde{q}\cdot D(\tilde{q})-\tilde{p}\cdot D(\tilde{p})]=\sum_{i\in N}[U_i(f^i(q,m_i))-U_i(f^i(p,m_i))].
\end{equation}
\end{cor}

That is, we can calculate the change in the sum of the utility in this trade using only the aggregated demand $D(\tilde{p})$. Therefore, in this economy, the actual information on $u_i$ is not necessary for welfare analysis, as in the usual partial equilibrium theory.\footnote{Note that, the producer's surplus corresponds to the total profit of producers, and, in the quasi-linear economy, an increase in profit means an increase in the consumption of the numeraire good by someone. Therefore, the total surplus coincides with the total increase in the sum of the utility in the quasi-linear economy according to the trade in this market.}

\subsection{Comparison with Related Work}
Most studies on the uniqueness of equilibrium do not use primitive assumptions on the economy itself, but rather make assumptions on what is derived from the economy. For example, Arrow et al. (1959) found that if the excess demand function is gross substitute, then the equilibrium price is unique up to normalization. Mas-Colell (1991) summarized several classical results in this area. He argued that the gross substitution of the excess demand function is no longer a sufficient condition for the uniqueness of the equilibrium price when production is introduced. Instead, his paper focused on the weak axiom of revealed preference for the excess demand function, and pointed out that the set of equilibrium prices is convex when this condition holds. If the economy is regular, then the set of normalized equilibrium prices becomes discrete, but any discrete convex set is a singleton. Thus, the uniqueness of the equilibrium price is obtained for this case. On the other hand, if the weak axiom of revealed preference is not satisfied, an economy with multiple equilibrium prices can easily be created by introducing production technology that is constant returns to scale. In this sense, Mas-Colell concluded that the weak axiom of revealed preference is approximately ``necessary and sufficient'' for the equilibrium price to be unique. This argument was discussed again in Chapter 17 of Mas-Colell et al. (1995).

However, both the gross substitution and the weak axiom of revealed preference on the excess demand function have a weakness that has been widely criticized in this context. Specifically, since the excess demand function is not a primitive component of the economy, but a function calculated from the primitive components of the economy, it is difficult to determine what assumptions are being made about the economy by placing assumptions on the excess demand function. Recently, Gim\'enez (2022) proved the uniqueness of the equilibrium price in a pure exchange economy with two commodities and two consumers, given some monotonicity assumption on the offer curves. However, because the offer curve is also not a primitive element of the economy, the same problem as above arises, and it becomes difficult to evaluate the strength of the assumption.

As already mentioned in the introduction, even in a two-commodity, two-consumer case, Exercise 15.B.6 of Mas-Colell et al. (1995) presents a numerical example of a pure exchange economy such that there are multiple equilibrium prices. In this economy, the utility functions of both consumers are CES type, and so this economy is not particularly strange. Therefore, the assumption for the uniqueness of the equilibrium price must be somewhat strong in the sense that it must rule out this example. The problem is to evaluate the strength of the assumption. However, it is difficult to evaluate the mathematical strength of assumptions on the excess demand function or the offer curves, because they are not primitives of the economy. Thus, perhaps these assumptions are ridiculously strong conditions. In fact, Debreu (1972) stated that the known sufficient conditions for the uniqueness of equilibrium price are ``exceedingly strong'', because, as noted above, the strength of the assumptions cannot be measured, and even a simple counterexample had been found.

On the other hand, this paper provides conditions for the uniqueness of the equilibrium price in an economy where there are $L$ commodities, $n$ consumers, and $\mu$ producers. Unlike the various conditions discussed above, our conditions are requirements on the utility functions and production sets, not on the excess demand function or the offer curve. This is a feature of the present paper in this context. The quasi-linear condition is itself a primitive condition for the economy, and thus avoids the above criticisms. Moreover, as mentioned at the end of section 2, the quasi-linear condition is independent of gross substitution. Therefore, as far as we can assess, the theorem in this paper is a good enough result in this context.

One may think that in Theorem 1, only the uniqueness of the equilibrium price is important, and its local stability is not so important. However, as the proof shows, local stability is in fact necessary for the proof of the uniqueness. It is incorporated into one important step of the proof for a completely different reason than in Arrow et al. (1959).

We would like to discuss this point in some detail. Suppose that $p^*$ is the equilibrium price of a quasi-linear economy, and the production function $y^j(p)$ is differentiable at this point. By an extension of Hotelling's lemma, we can show that $y^j(p)$ is in fact a gradient vector field of the profit function $\pi^j(p)$. Since the profit function is convex, we have that $Dy^j(p)$ is positive semi-definite. Moreover, for a quasi-linear economy, we can show by a direct calculation that
\begin{equation}\label{EXCESS}
DX(p^*)=\sum_{i\in N}S_{f^i}(p^*,m^i(p^*)),
\end{equation}
where $p^*$ is an equilibrium price, and $X$ is the sum of the excess demand functions for each consumer. From these two facts, we can show that $D\zeta(p^*)$ is negative definite on the space of all vectors normal to $p^*$. Therefore, the index of $p^*$ is $+1$. Because any equilibrium price has a non-zero index, this economy is regular, and the Poincar\'e--Hopf index theorem states that the sum of the index of normalized equilibrium prices is $+1$. Because the index of any equilibrium price is $+1$, this indicates that such a normalized equilibrium price is unique.

The above argument can apply only when $y^j$ is differentiable at $p^*$. Actually, our Assumption P is too weak and can only prove the continuity of $y^j$. Hence, we use approximation by using {\bf mollifier}, which is frequently used in Fourier analysis. As is well known, when a convex function is approximated by a convolution with a mollifier, the approximate function becomes a smooth convex function. Using this approximation for the profit function $\pi^j(p)$, we can obtain the `approximated' profit function, and using this function, we construct a smooth approximation of the excess demand function whose derivative is negative definite on the space of all vectors normal to $p^*$. Applying the above arguments, we obtain the desired result.

However, in order to perform this approximation properly, it must first be shown that the set of normalized equilibrium prices is discrete. Local stability is crucial in demonstrating this fact. If every equilibrium price is locally stable, the set of normalized equilibrium prices is discrete. Therefore, we first show that every equilibrium price is locally stable, and then use the above logic to derive the result. This is why local stability is necessary for the derivation of our result.

So, why are all equilibrium prices locally stable in a quasi-linear economy? The answer is obtained by the theory of {\bf no-trade equilibria}. Balasko (1978, Theorem 1) showed that in a pure exchange economy, any no-trade equilibrium price is locally stable. This result was in fact substantially shown in Kihlstrom et al. (1976, Lemma 1). Namely, they showed that if the initial endowments become the equilibrium allocation, then (\ref{EXCESS}) holds for the corresponding equilibrium price. If the economy is not quasi-linear, (\ref{EXCESS}) may not hold at some equilibrium price because the {\bf income effect} that arises from the gap between the initial endowments and the equilibrium allocation has a non-negligible power. In a quasi-linear economy, however, the income effect affects only the numeraire good, and when we aggregate the excess demand function of each consumer, the error with (\ref{EXCESS}) is equal to the value of the excess demand function divided by $p_L$ (see Lemma \ref{LEM6} and (\ref{Slutsky}) in Step 2 of the proof of Theorem \ref{THM1}). Thus, this effect is canceled out when the price is an equilibrium price. As a result, the result that holds at the no-trade equilibrium price is restored at any equilibrium price.

Hayashi (2017) provided details on studies discussing the relationship between partial and general equilibria. In this connection, let us evaluate our assumption within the partial equilibrium framework. We find that Assumption P is a sufficiently weak. This assumption prohibits the case where the supply curve is horizontal but prohibits almost nothing else. For example, the supply curve may diverge at a finite output level, or, conversely, may rise only to a finite price-level even if the output diverges. In contrast, the assumptions for consumers are somewhat strong. Actually, Assumption Q implies that the range of the demand curve $D(x)$ must be $\mathbb{R}_{++}$. Although this assumption is strong, we cannot remove this in producing general results. For some special cases, weaker assumptions could be sufficient.

Finally, the study discussed in section 3.2 is related to section 4 of Osana (1992). This paper, written in Japanese, discusses not only consumer surplus but also the relationship between this and equivalent and compensating variations. The relationship between Stokes' theorem and these results is also discussed.

\section{Concluding Remarks}
We have shown that in a quasi-linear economy, the equilibrium price is uniquely determined and is locally stable. Compared with similar results, a feature of this result is that there is no assumption imposed on the excess demand function. Moreover, we have exhibited that in this economy, consumers' surplus can be defined, and coincides with the change in the sum of utilities.

There are a few future tasks. First, we assumed that $u$ is twice continuously differentiable and the Hessian matrix is negative definite. This is needed for assuring the differentiability of the demand function. However, it can be shown by using some approximation techniques that our results still hold for non-smooth demands.

Second, we prohibited the case in which the boundary of the production set includes a line outside $-\mathbb{R}^L_+$. This assumption is needed to prevent the multi-valuedness of $y^i(p)$. However, for some accurate approximation techniques, we can overcome this technical difficulty. In particular, we may be able to show the same result for economies with technologies of constant returns to scale.

\section{Proofs}

\subsection{Lemmas}
In this subsection, we show several lemmas.

First, suppose that $E$ is a quasi-linear economy and choose any $s\ge 0$. Let
\[\Omega_i^s=\Omega_i\cap (\mathbb{R}^{L-1}_+\times [-s,+\infty[),\]
\[Y_j^s=Y_j\cap \{y\in \mathbb{R}^L|\|y\|\le s\}.\]
Consider the modified economy
\[E_s=(N,M,(\Omega_i^s)_{i\in N}, (U_i)_{i\in N}, (\omega^i)_{i\in N},(Y_j^s)_{j\in M},(\theta_{ij})_{i\in N,j\in M}).\]
Let $f^i_s$ be the demand function of consumer $i$ in economy $E_s$, $y^j_s, \pi^j_s$ be the supply function and the profit function of producer $j$, respectively, in economy $E_s$, $X^i_s$ be the excess demand function of consumer $i$ in the economy $E_s$, $X_s=\sum_{i\in N}X^i_s$, and $\zeta_s$ be the excess demand function of economy $E_s$.

\begin{lem}\label{LEM1}
Suppose that $E$ is a quasi-linear economy. Then, there exists a continuously differentiable function $\tilde{x}^i:\mathbb{R}^{L-1}_{++}\to \mathbb{R}^{L-1}_{++}$ such that
\[\tilde{y}=\tilde{x}^i(\tilde{p})\Leftrightarrow \nabla u_i(\tilde{y})=\tilde{p}.\]
\end{lem}

\begin{proof}
Let $\tilde{p}\in \mathbb{R}^{L-1}_{++}$. By Assumption Q, there exists a solution $\tilde{x}^*\in\mathbb{R}^{L-1}_{++}$ of the following equation:
\begin{equation}\label{EQ}
\nabla u_i(\tilde{x})=\tilde{p}.
\end{equation}
Consider the following optimization problem:
\begin{align*}
\max~~~~~&~u_i(\tilde{x})-\tilde{p}\cdot \tilde{x}\\
\mbox{subject to. }&~\tilde{x}\in \mathbb{R}^{L-1}_{++}.
\end{align*}
Because of either Assumption F or Assumption S1, we have that $u_i$ is strictly concave on $\mathbb{R}^{L-1}_{++}$, and thus 1) any solution to (\ref{EQ}) is also a solution to this problem, and 2) the solution to the above problem is unique. Therefore, $\tilde{x}^*$ is the unique solution to the equation (\ref{EQ}), and thus we can define $\tilde{x}^i(\tilde{p})=\tilde{x}^*$. Because $D^2u_i(\tilde{x}^*)$ is negative definite, it is regular, and thus by the inverse function theorem, we have that $\tilde{x}^i(\tilde{p})$ is continuously differentiable. This completes the proof. $\blacksquare$
\end{proof}

For any $\tilde{p}\in \mathbb{R}^{L-1}_{++}$ and $m\in \mathbb{R}_{++}$, let
\begin{equation}\label{QLL}
x_L^i(\tilde{p},m)=m-\tilde{p}\cdot \tilde{x}^i(\tilde{p}).
\end{equation}

\begin{lem}\label{LEM2}
Suppose that $E$ is a first-type quasi-linear economy and $s\ge 0$, and choose any $(p,m)\in\mathbb{R}^L_{++}\times \mathbb{R}_{++}$. Define $(q,w)=p_L^{-1}(p,m)$. Then,\footnote{From this result, in the first-type quasi-linear economy, we can consider that the domain of $f^i$ is $\mathbb{R}^L_{++}\times \mathbb{R}$. This fact is used in the proof of Lemma \ref{LEM7}.}
\[f^i(p,m)=(\tilde{x}^i(\tilde{q}),x_L^i(\tilde{q},w)),\]
and if $x_L^i(\tilde{q},w)\ge -s$, then
\[f^i_s(p,m)=(\tilde{x}^i(\tilde{q}),x_L^i(\tilde{q},w)).\]
\end{lem}

\begin{proof}
By Assumption F, we have that $u_i(\tilde{x})$ is strictly concave on $\mathbb{R}^{L-1}_{++}$. Note that, because $u_i$ is concave on $\mathbb{R}^{L-1}_+$, $U_i$ is also concave on $\Omega_i$.

Define
\[x^*=(\tilde{x}^i(\tilde{q}),x_L^i(\tilde{q},w)).\]
First, we show the latter claim of this lemma. Suppose that $x_L^*\ge -s$. Then, we have that $x^*\in \Omega_i^s$. Moreover,
\[p\cdot x^*=p_L(\tilde{q}\cdot \tilde{x}^i(\tilde{q})+x_L^i(\tilde{q},w))=p_Lw=m.\]
By Lagrange's multiplier rule, we have that $x^*\in f^i_s(p,m)$. Suppose that $y^*\in f^i_s(p,m)$ and $x^*\neq y^*$. Then, $U_i(x^*)=U_i(y^*)$. If $\tilde{x}^*=\tilde{y}^*$, then $x_L^*=y_L^*$ by equation $U_i(x^*)=U_i(y^*)$, which contradicts $x^*\neq y^*$. Thus, we have that $\tilde{x}^*\neq \tilde{y}^*$. Define $z(t)=(1-t)x^*+ty^*$. Then, for every $t\in [0,1]$, we have that $p\cdot z(t)\le m$, and thus $U_i(z(t))\le U_i(x^*)$. Because $U_i$ is concave, we have that $U_i(z(t))=U_i(x^*)$ for all $t\in [0,1]$. Set $t_1=\frac{1}{2}$ and $t_2=\frac{1}{4}$, and let $z^*=z(t_1), z^+=z(t_2)$. Then, $\tilde{z}^*\in \mathbb{R}^{L-1}_{++}$, $p\cdot z^*\le m$, and $U_i(z^*)=U_i(x^*)$. However, because $u_i$ is strictly concave on $\mathbb{R}^{L-1}_{++}$, we have that 
\[u_i(\tilde{z}^+)>\frac{1}{2}u_i(\tilde{x}^*)+\frac{1}{2}u_i(\tilde{z}^*),\]
which implies that $U_i(z^+)>U_i(x^*)$. This contradicts $U_i(z^+)=U_i(x^*)$. Therefore, $f^i_s(p,m)=\{x^*\}$, as desired.

Next, we show the former claim of this lemma. Clearly $x^*\in \Omega_i$. Again by Lagrange's multiplier rule, we have that $x^*\in f^i(p,m)$. If $y^*\in f^i(p,m)$ and $x^*\neq y^*$, then choose $s>0$ so large that $x^*,y^*\in \Omega_i^s$. Then, $y^*\in f^i_s(p,m)$, which is a contradiction. This completes the proof. $\blacksquare$
\end{proof}

\begin{lem}\label{LEM3}
Suppose that $E$ is a first-type quasi-linear economy, and $s\ge 0$. Choose any $(p,m)\in \mathbb{R}^L_{++}\times \mathbb{R}_{++}$, and define
\[x^*=(\tilde{x}^i(p_L^{-1}\tilde{p}),x_L^i(p_L^{-1}\tilde{p},p_L^{-1}m)).\]
Suppose that $x^*_L<-s$. Let $\bar{m}=m+p_Ls$, and $\tilde{x}^+\in \mathbb{R}^{L-1}_{++}$ be a solution to the problem
\begin{align*}
\max~~~~~&~u_i(\tilde{x})\\
\mbox{subject to. }&~\tilde{x}\in \mathbb{R}^{L-1}_+,\\
&~\tilde{p}\cdot \tilde{x}\le \bar{m},
\end{align*}
and define $x_L^+=-s$. Then, $f^i_s(p,m)=(\tilde{x}^+,x_L^+)$.\footnote{Note that, the existence of such an $\tilde{x}^+$ is assumed in Assumption Q.}
\end{lem}

\begin{proof}
First, because of Assumption F, $u_i$ is strictly concave on $\mathbb{R}^{L-1}_{++}$, and thus the solution to the above problem is unique. Therefore, if $\tilde{y}\in \mathbb{R}^{L-1}_+$ and $\tilde{p}\cdot \tilde{y}\le \bar{m}$, then either $\tilde{y}=\tilde{x}^+$ or $u_i(\tilde{y})<u_i(\tilde{x}^+)$. Because $u_i$ is increasing on $\mathbb{R}^{L-1}_{++}$, we have that $\tilde{p}\cdot \tilde{x}^+=\bar{m}$. Because $f^i(p,m)=x^*$, we have that $U_i(x^*)>U_i(x^+)$.

Suppose that there exists $y^+\in \Omega_i^s$ such that $x^+\neq y^+$, $p\cdot y^+\le m$ and $U_i(y^+)\ge U_i(x^+)$. Then, $\tilde{p}\cdot \tilde{y}^+\le \bar{m}$. If $y_L^+=-s$, then we have that $\tilde{y}^+\neq \tilde{x}^+$, and thus $u_i(\tilde{y}^+)<u_i(\tilde{x}^+)$, which is a contradiction. Therefore, $y_L^+>-s$, and there exists $t\in ]0,1[$ such that for $z^+=(1-t)x^*+ty^+$, $z_L^+=-s$. Because $u_i$ is concave on $\mathbb{R}^{L-1}_+$, we have that $u_i(\tilde{z}^+)\ge (1-t)u_i(\tilde{x}^*)+tu_i(\tilde{y}^+)$, which implies that $U_i(z^+)>U_i(x^+)$, and thus $u_i(\tilde{z}^+)>u_i(\tilde{x}^+)$. However, $\tilde{p}\cdot \tilde{z}^+\le m+p_Ls=\bar{m}$, which contradicts the definition of $\tilde{x}^+$. Thus, we have that $f^i_s(p,m)=\{x^+\}$, as desired. This completes the proof. $\blacksquare$
\end{proof}

\begin{lem}\label{LEM4}
Suppose that $E$ is a second-type quasi-linear economy, and $\hat{E}$ is the first-type quasi-linear economy such that all components except for $\Omega_i$ are the same as $E$. Then, the demand function $f^i$ in $E$ coincides with $f^i_0$, where $f^i_0$ is the demand function in $\hat{E}_0$.
\end{lem}

\begin{proof}
Obvious by definition. $\blacksquare$
\end{proof}

\begin{lem}\label{LEM5}
Suppose that $E$ is a quasi-linear economy. Then, for every $i\in N$, $f^i$ is a single-valued continuous function. Moreover, if $x=f^i(p,m)$, then $\tilde{x}\in \mathbb{R}^{L-1}_{++}$. Furthermore, Walras' law
\begin{equation}\label{WL}
p\cdot f^i(p,m)=m
\end{equation}
and homogeneity of degree zero
\begin{equation}\label{HDZ}
f^i(ap,am)=f^i(p,m)\mbox{ for all }a>0
\end{equation}
hold.
\end{lem}

\begin{proof}
If $E$ is first-type, then by Lemma \ref{LEM2}, we have that $f^i$ is single-valued and continuously differentiable, and $\tilde{f}^i(p,m)\in \mathbb{R}^{L-1}_{++}$. If $E$ is second-type, then by Lemmas \ref{LEM2}-\ref{LEM4}, $f^i$ is single-valued and $\tilde{f}^i(p,m)\in \mathbb{R}^{L-1}_{++}$, and by Berge's maximum theorem, $f^i$ is continuous.

It is clear that $f^i$ is homogeneous of degree zero. Because $U_i$ is locally non-satiated, we have that $f^i$ satisfies Walras' law. This completes the proof. $\blacksquare$
\end{proof}

\begin{lem}\label{LEM6}
Suppose that $E$ is a quasi-linear economy. If either $E$ is first-type or $f^i_L(p,m)>0$, then $f^i$ is continuously differentiable at $(p,m)$, and
\begin{equation}
\frac{\partial f^i_{\ell}}{\partial m}(p,m)=\begin{cases}
0 & \mbox{if }1\le \ell\le L-1,\\
\frac{1}{p_L} & \mbox{if }\ell=L.
\end{cases}\label{QLD}
\end{equation}
\end{lem}

\begin{proof}
Let $\tilde{\Omega}_i$ denote the interior of $\Omega_i$. Suppose that $x=f^i(p,m)$ and either the economy is first-type or $x_L>0$. Then, $x\in \tilde{\Omega}_i$, and because of Lemmas \ref{LEM2} and \ref{LEM4}, there exists an open neighborhood $V$ of $(p,m)$ such that if $(q,w)\in V$, then
\[f^i(q,w)=(\tilde{x}^i(q_L^{-1}\tilde{q}),x_L^i(q_L^{-1}\tilde{q},q_L^{-1}w)),\]
where $x_L^i$ is defined in (\ref{QLL}). Therefore, $f^i$ is continuously differentiable on $V$ and (\ref{QLD}) holds. This completes the proof. $\blacksquare$
\end{proof}

\begin{lem}\label{LEM7}
Suppose that $E$ is a quasi-linear economy, and fix $(p,m)\in \mathbb{R}^L_{++}\times \mathbb{R}_{++}$. Suppose also that either $E$ is first-type or $f^i_L(p,m)>0$. Then, the Slutsky matrix $S_{f^i}(p,m)$ satisfies the following three properties.
\begin{description}
\item{(R)} The rank of $S_{f^i}(p,m)$ is $L-1$. Moreover, $p^TS_{f^i}(p,m)=0^T$ and $S_{f^i}(p,m)p=0$.

\item{(ND)} For every $v\in \mathbb{R}^L$ such that $v\neq 0$ and $p\cdot v=0$, $v^TS_{f^i}(p,m)v<0$.

\item{(S)} The matrix $S_{f^i}(p,m)$ is symmetric.
\end{description}
\end{lem}

\begin{proof}
First, choose an open neighborhood $U$ of $(p,m)$ such that $f^i$ is continuously differentiable on $U$. Because of Lemma \ref{LEM5}, we have that
\[q\cdot f^i(q,w)=w\]
for every $(q,w)\in U$ and
\[f^i(ap,am)=f^i(p,m)\]
for every $a>0$. Hence, by differentiation,
\[f^i_j(p,m)+\sum_{k=1}^Lp_k\frac{\partial f^i_k}{\partial p_j}(p,m)=0,\]
\[p\cdot \frac{\partial f^i}{\partial m}(p,m)=1,\]
\[\sum_{j=1}^Lp_j\frac{\partial f^i_k}{\partial p_j}(p,m)+m\frac{\partial f^i_k}{\partial m}(p,m)=0,\]
and thus we have
\[p^TS_{f^i}(p,m)=0^T,\ S_{f^i}(p,m)p=0.\]
Second, for $x=f^i(p,m)$, define 
\[E^x_i(q)=\inf\{q\cdot y|U_i(y)\ge U_i(x)\}.\]
If the economy $E$ is second-type, we can apply Lemma 1 of Hosoya (2020) directly, and obtain that $E^x_i$ is concave and positive, and the following equation (Shephard's lemma) holds:
\begin{equation}\label{Shephard}
\nabla E^x_i(q)=f^i(q,E^x_i(q)),\ E^x_i(p)=m.
\end{equation}
Suppose that the economy is first-type. We show that for every $q\in \mathbb{R}^L_{++}$, $E_i^x(q)$ is finite. Define $r=\frac{1}{q_L}q$, and
\[x^*=(\tilde{x}^i(\tilde{r}),U_i(x)-u_i(\tilde{x}^i(\tilde{r}))).\]
Then, $U_i(x^*)=U_i(x)$. Let $q\cdot x^*=w$. By Lemma \ref{LEM2}, $x^*=f^i(q,w)$. Therefore, if $q\cdot y\le w$ and $y\neq x^*$, then $U_i(y)<U_i(x)$. This implies that $E^x_i(q)=w>-\infty$. In particular, if $q=p$, then $x^*=x$, and thus $E^x_i(p)=m$.

If $q^1,q^2\in \mathbb{R}^L_{++}$ and $0\le t\le 1$, then $U_i(y)\ge U_i(x)$ implies that
\[[(1-t)q^1+tq^2]\cdot y=(1-t)q^1\cdot y+tq^2\cdot y\ge (1-t)E_i^x(q^1)+tE_i^x(q^2),\]
and thus, we have that
\[E_i^x((1-t)q^1+tq^2)\ge (1-t)E_i^x(q^1)+tE_i^x(q^2).\]
This implies that $E_i^x$ is concave. Because every concave function defined on $\mathbb{R}^L_{++}$ is continuous, we have that $E^x$ is continuous. Because $E^x_i(p)=m>0$, there exists a neighborhood $U$ of $p$ such that $E^x_i(q)>0$ for all $q\in U$. By the same proof as in that of Lemma 1 of Hosoya (2020), again we can show that (\ref{Shephard}) holds on $U$. Therefore, under our assumtions, we have that (\ref{Shephard}) holds on some neighborhood of $p$. Because $f^i$ is continuously differentiable around $(p,m)$, to differentiate both sides of (\ref{Shephard}), we have that
\[D^2E^x_i(p)=S_{f^i}(p,m).\]
Therefore, (S) holds because of Young's theorem, and $S_{f^i}(p,m)$ is negative semi-definite.

Third, let $\tilde{\Omega}_i$ denote the interior of $\Omega_i$, and for $y\in \tilde{\Omega}_i$, define
\[g^i(y)=\nabla U_i(y),\]
and for $j,k\in \{1,...,L-1\}$, define
\[a^i_{jk}(y)=\frac{\partial g^i_j}{\partial x_k}(y)-\frac{\partial g^i_j}{\partial x_n}(y)g^i_k(y).\]
The $(L-1)\times (L-1)$ matrix-valued function $A_{g^i}(y)=(a^i_{jk}(y))_{j,k=1}^{L-1}$ is called the {\bf Antonelli matrix}. Samuelson (1950) showed that if $y=f^i(q,w)\in \tilde{\Omega}_i$, then $A_{g^i}(y)$ is regular, and the inverse matrix of $A_{g^i}(y)$ coincides with $(s_{jk}^i(q,w))_{j,k=1}^{L-1}$. By the assumption of this lemma, we have that if $x=f^i(p,m)$, then $x\in \tilde{\Omega}_i$, and thus, (R) holds. Moreover, because $S_{f^i}(p,m)$ is symmetric, there exists an orthogonal matrix $P$ such that
\[S_{f^i}(p,m)=P^T\begin{pmatrix}
\lambda_1 & 0 & ... & 0\\
0 & \lambda_2 & ... & 0\\
\vdots & \vdots & \ddots & \vdots\\
0 & 0 & ... & \lambda_L
\end{pmatrix}P,\]
where each $\lambda_j\le 0$ is an eigenvalue of $S_{f^i}(p,m)$. Because of (R), we have that there exists exactly one $j$ such that $\lambda_j=0$, and $\lambda_k<0$ whenever $k\neq j$. This implies that (ND) holds, which completes the proof. $\blacksquare$
\end{proof}

\begin{lem}\label{LEM8}
Suppose that $E$ is a quasi-linear economy. If $y^j(p)$ is nonempty, then it is a singleton. The same holds for $y^j_s(p)$ when $s>0$. Moreover, $y^j_s$ is a single-valued continuous function defined on $\mathbb{R}^L_{++}$, and the profit function $\pi^j_s$ is convex and satisfies $\nabla \pi^j_s(p)=y^j_s(p)$ for all $p\in \mathbb{R}^L_{++}$.\footnote{This is a variety of Hotelling's lemma.} If, in addition, $\|y^j_s(p)\|<s$, then $y^j(q)=y^j_s(q)$ and $\pi^j(q)=\pi^j_s(q)$ on an open neighborhood of $p$.
\end{lem}

\begin{proof}
First, suppose that $y^1,y^2\in y^j(p)$ and $y^1\neq y^2$. Define $y^3=\frac{1}{2}(y^1+y^2)$. Because $p\cdot y^3=p\cdot y^1+p\cdot y^2$, we have that $y^3\in y^j(p)$. Because $0\in Y_j$, $p\cdot y^1=p\cdot y^2\ge 0$, and thus either $y^i=0$ or $y^i\notin -\mathbb{R}^L_+$ for $i\in \{1,2\}$. If $y^1\neq 0$ and $y^2\neq 0$, then $y^3$ is in the interior of $Y_j$. This implies that $y^3\notin y^j(p)$, which is a contradiction. Therefore, either $y^1=0$ or $y^2=0$. We assume that $y^1=0$, and define $y^4=\frac{1}{2}(y^3+y^2)$. Then, $y^4\in y^j(p)$. Because $y^2\notin -\mathbb{R}^L_+$, we have that $y^3\notin -\mathbb{R}^L_+$, and thus $y^4$ is in the interior of $Y_j$. This implies that $y^4\notin y^j(p)$, which is a contradiction. Therefore, $y^j(p)$ is either a singleton or the empty set.

The same proof can be applied for $y^j_s(p)$.\footnote{Note that, if $\|y^1\|=\|y^2\|=s$, then $\|y^3\|<s$.} Because $Y_j^s$ is compact, we have that $y^j_s(p)$ is always single-valued. By Berge's maximum theorem, we have that $y^j_s(p)$ is continuous.

Choose any $p,q\in \mathbb{R}^L_{++}$ and let $0\le t\le 1$ and $r=(1-t)p+tq$. Then,
\[r\cdot y^j_s(r)=(1-t)p\cdot y^j_s(r)+tq\cdot y^j_s(r)\le (1-t)p\cdot y^j_s(p)+tq\cdot y^j_s(q),\]
which implies that
\[\pi^j_s(r)\le (1-t)\pi^j_s(p)+t\pi^j_s(q).\]
Therefore, $\pi^j_s$ is a convex function. Recall that $e_{\ell}$ denotes the $\ell$-th unit vector, and let $q=p+he_{\ell}$. Then, $p\cdot y^j_s(q)\le p\cdot y^j_s(p)$, and thus
\begin{align*}
\pi^j_s(q)-\pi^j_s(p)=&~q\cdot y^j_s(q)-p\cdot y^j_s(p)\\
=&~p\cdot (y^j_s(q)-y^j_s(p))+hy^j_{s,\ell}(q)\\
\le&~hy^j_{s,\ell}(q).
\end{align*}
This implies that
\[\lim_{h\uparrow 0}\frac{\pi^j_s(q)-\pi^j_s(p)}{h}\ge y^j_{s,\ell}(p)\ge \lim_{h\downarrow 0}\frac{\pi^j_s(q)-\pi^j_s(p)}{h}.\]
Because $\pi^j_s$ is convex, the above inequalities turn into equalities, and thus
\[\nabla \pi^j_s(p)=y^j_s(p),\]
as desired.

Finally, suppose that $\|y^j_s(p)\|<s$. Because the function $y^j_s$ is continuous, there exists an open neighborhood $U$ of $p$ such that $\|y^j_s(q)\|<s$ for any $q\in U$. Therefore, it suffices to show that $\|y^j_s(p)\|<s$ implies that $y^j(p)=y^j_s(p)$. Hence, choose any $y\in Y_j$, and suppose that $p\cdot y>p\cdot y^j_s(p)$. Let $y(t)=(1-t)y^j_s(p)+ty$. Then, $p\cdot y(t)>p\cdot y^j_s(p)$ for all $t>0$, and if $t$ is sufficiently small, then $y(t)\in Y_j$ and $\|y(t)\|<s$, which contradicts the definition of $y^j_s(p)$. Thus, $y^j_s(p)=y^j(p)$. This completes the proof. $\blacksquare$
\end{proof}

\begin{lem}\label{LEM9}
Suppose that $E$ is an economy such that Assumption P holds. Then, there exists $p^+\in \mathbb{R}^L_{++}$ such that, for any $j\in J$, $y^j(p^+)$ is nonempty.
\end{lem}

\begin{proof}
Construct another economy $E'=(N', M, (\Omega_i')_{i\in N'}, (U_i')_{i\in N'}, (\omega^i)_{i\in N'}$, $(Y_j)_{j\in M}, (\theta_{ij}')_{i\in N',j\in M})$ such that, 1) $N'=\{1,2\}$, 2) $\Omega_1'=\Omega_2'=\mathbb{R}^L_+$, 3) $U_1'(x)=U_2'(x)=(x_1...x_L)^{1/L}$, 4) $\omega^1=\omega^2=(1,1,...,1)$, and 5) $\theta_{ij}'=\frac{1}{2}$ for all $i\in N',j\in M$. It is known that this economy has at least one equilibrium price $p^+\in \mathbb{R}^L_{++}$.\footnote{See, for example, Theorem I of Arrow and Debreu (1954), or section 5.7 of Debreu (1959).} Because of the definition of the equilibrium price, we have that $y^j(p^+)$ is nonempty for each $j\in M$. This completes the proof. $\blacksquare$
\end{proof}

\begin{lem}\label{LEM10}
Suppose that $E$ is a quasi-linear economy. Define 
\[B_{\omega}=\{(x,y)\in \bar{A}_{\omega}|U_i(x^i)\ge U_i(\omega^i)\mbox{ for each }i\in N\}.\]
Then, $B_{\omega}$ is compact.\footnote{We want to use this lemma to prove Proposition \ref{PROP1}. Thus, in the proof, we admit the case in which $\omega^i_L=0$ for all $i\in N$.}
\end{lem}

\begin{proof}
If $E$ is a second-type quasi-linear economy, then this property can be shown by the usual arguments.\footnote{See, for example, subsection 3.3.1 of Arrow and Debreu (1954).} Hence, we assume that $E$ is a first-type quasi-linear economy.

Suppose that $B_{\omega}$ is not compact. Because $B_{\omega}$ is closed, this implies that $B_{\omega}$ is unbounded. Therefore, there exists a sequence $(x^k,y^k)$ in $B_{\omega}$ such that $\|(x^k,y^k)\|\to \infty$ as $k\to \infty$.

First, suppose that there exists $s\ge 0$ such that $x^{ik}_L\ge -s$ for all $i$ and all sufficiently large $k$. Then, we can assume that for all $k$, $(x^k,y^k)$ is admissible in economy $E_s$. Thus, by the same argument as in subsection 3.3.1 of Arrow and Debreu (1954), we can show that $(x^k,y^k)$ is bounded, which is a contradiction. Therefore, we can assume that there exists $i\in N$ such that $x^{ik}_L$ is unbounded from below. Let $I$ be the set of such $i$, and by taking a subsequence, we assume that, for all $i\in I$, $x^{ik}_L\to -\infty$ as $k\to \infty$. Define $y^{k*}=\sum_{j\in M}y^{jk}$, and
\[\tilde{z}^k=\sum_{i\in N}\tilde{\omega}^i+\tilde{y}^{k*}.\]
Then, $u_i(\tilde{x}^{ik})\le u_i(\tilde{z}^k)$. Choose $p^+$ in Lemma \ref{LEM9}. Then,
\[p^+\cdot y^{k*}\le \sum_{j\in M}\pi^j(p^+)\equiv m^*,\]
which implies that
\[\tilde{p}^+\cdot \tilde{y}^{k*}\le m^*-p^+_Ly^{k*}_L.\]
For $i\in N\setminus I$, define $w^i=\inf_kx^{ik}_L$. Moreover, define
\[x^{k*}_L=\sum_{i\in I}x^{ik}_L,\ w^*=\sum_{i\in N}\omega^i_L-\sum_{i\in N\setminus I}w^i.\]
Because
\[\sum_{i\in N}x^{ik}_L=\sum_{i\in N}\omega^i_L+y^{k*}_L,\]
we have that
\[x^{k*}_L-w^*\le y^{k*}_L.\]
Now, choose any $i\in I$, and let $g^i(\tilde{p},m)$ be the solution to (\ref{QLU2}). Then, by the above arguments,
\[u_i(\tilde{z}^k)\le u_i(g^i(\tilde{p}^+,m^+(p^+_Lx^{k*}_L))),\]
where
\[m^+(a)=m^*+\sum_{i\in N}\tilde{p}^+\cdot \tilde{\omega}^i+p_L^+w^*-a.\]
Define $c_i(m)=u_i(g^i(\tilde{p}^+,m))$. Because (\ref{QLU2}) has an inner solution, we have that $c_i$ is continuously differentiable. Moreover, by the envelope theorem, we have that $c_i'(m)=\frac{\frac{\partial u_i}{\partial x_1}(g^i(\tilde{p}^+,m))}{p_1^+}$. Choose any $t\in [0,1]$ and $m_1,m_2>0$ with $m_1<m_2$, and let $m_3=(1-t)m_1+tm_2$. Because $u_i$ is concave, we have that
\[c_i(m_3)\ge u_i((1-t)g^i(\tilde{p}^+,m_1)+tg^i(\tilde{p}^+,m_2))\ge (1-t)c_i(m_1)+tc_i(m_2),\]
which implies that $c_i$ is concave. By Assumption Q, there exists $\tilde{x}^{i*}\in \mathbb{R}^{L-1}_{++}$ such that $\nabla u_i(\tilde{x}^{i*})=\frac{1}{2|I|p_L^+}\tilde{p}^+$, where $|I|$ denotes the number of elements included in $I$. Note that, because of Lagrange's multiplier rule, $\tilde{x}^{i*}=g^i(\tilde{p}^+,\tilde{p}^+\cdot \tilde{x}^{i*})$. If $m_0=\max_{i\in I}\tilde{p}^+\cdot \tilde{x}^{i*}$, then we have that $c_i'(m)\le \frac{1}{2|I|p_L^+}$ for all $i\in I$ and $m\ge m_0$. Because $x^{k*}_L\to -\infty$, we have that $m^+(p_L^+x^{k*}_L)\ge m_0$ for any sufficiently large $k$, and thus
\begin{align*}
\sum_{i\in I}U_i(x^{ik})=&~\sum_{i\in I}u_i(\tilde{x}^{ik})+x^{k*}_L\\
\le&~\sum_{i\in I}u_i(\tilde{z}^k)+x^{k*}_L\\
\le&~\sum_{i\in I}c_i(m^+(p^+_Lx^{k*}_L))+x^{k*}_L\to -\infty
\end{align*}
as $k\to \infty$, which contradicts the definition of $B_{\omega}$. This completes the proof. $\blacksquare$
\end{proof}

\begin{lem}\label{LEM11}
Suppose that $E$ is a quasi-linear economy and choose any $s>0$. Then, $\zeta_s$ is a single-valued function that satisfies the following.
\begin{enumerate}[1)]
\item $\zeta_s$ is continuous and satisfies (\ref{WL2}) and (\ref{HDZ2}).

\item There exists $S>0$ such that $\zeta_{s,\ell}(p)>-S$ for every $p\in \mathbb{R}^L_{++}$ and $\ell\in \{1,...,L\}$.

\item If $(p^k)$ is a sequence in $\mathbb{R}^L_{++}$ such that $p^k\to p\neq 0$ as $k\to 0$ and the set $J=\{\ell|p_{\ell}=0\}$ is nonempty, then $\|\zeta_s(p^k)\|\to \infty$ as $k\to \infty$.
\end{enumerate}
\end{lem}

\begin{proof}
Recall again that $e_{\ell}$ denotes the $\ell$-th unit vector. First, we treat a first-type quasi-linear economy. By Lemmas \ref{LEM2}-\ref{LEM3}, we have that $f^i_s$ is a single-valued function. Because of Berge's maximum theorem, we have that $f^i_s$ is continuous. Moreover, by Lemma \ref{LEM8}, $y^j_s$ is also single-valued and continuous. Therefore, $\zeta_s$ is single-valued and continuous.

It is easy to prove that $\zeta_s$ satisfies (\ref{WL2}) and (\ref{HDZ2}), and thus we omit the proof of this fact. Thus, $\zeta_s$ satisfies 1).

Because $\zeta_s(p)\gg -\sum_{i\in N}\omega^i-((\mu+1)s+1,(\mu+1)s+1,...,(\mu+1)s+1)$ for all $p\in \mathbb{R}^L_{++}$, we have that $\zeta_s$ satisfies 2).

Therefore, it suffices to show that 3) holds for $\zeta_s$. Suppose that $(p^k)$ is a sequence in $\mathbb{R}^L_{++}$ such that $p^k\to p\neq 0$ as $k\to 0$ and the set $J=\{\ell|p_{\ell}=0\}$ is nonempty, but $\|\zeta_s(p^k)\|\not\to\infty$. By taking a subsequence, we can assume that $\zeta_s(p^k)\to x$ as $k\to \infty$ for some $x\in \mathbb{R}^L$. Define
\[z^k=\zeta_s(p^k),\ y^{jk}=y^j_s(p^k),\]
\[x^{ik}=f^i_s(p^k,p^k\cdot \omega^i+\sum_{j\in M}\theta_{ij}p^k\cdot y^{jk}).\]
Because $(y^{jk})$ is bounded for each $j\in M$, we must have that $(x^{ik})$ is also bounded for each $i\in N$, and thus we can assume that $x^{ik}\to x^i$ and $y^{jk}\to y^j$ as $k\to \infty$. Note that $p^k\cdot y^{jk}\ge 0$, and thus $p\cdot y^j\ge 0$.

Suppose that $p_L=0$. Because $p\cdot x^i=p\cdot \omega^i+\sum_{j\in M}\theta_{ij}p\cdot y^j$ and $\sum_{i\in N}\omega^i\gg 0$, we have that there exists $i$ such that $x^i_{\ell}>0$ for some $\ell$ with $p_{\ell}>0$. Define $v^i=x^i-\varepsilon e_{\ell}+e_L$, where $\varepsilon>0$ is sufficiently small that $U_i(v^i)>U_i(x^i)$. Then, $U_i(v^i)>U_i(x^{ik})$ and $p^k\cdot v^i<p^k\cdot x^{ik}$ for some $k$, which is a contradiction.

Therefore, we have that $p_L>0$. Next, suppose that for some $i$, $u_i(\tilde{x}^i)=u_i(0)$. Because $U_i(x^i)\ge U_i(\omega^i)$, we have that $x^i_L\ge 0>-s$. Fix any $M^+>2\frac{\|\tilde{p}\|}{p_L}$. By Assumption Q, there exists $\tilde{x}^+\in \mathbb{R}^{L-1}_{++}$ such that $\frac{\partial u_i}{\partial x_j}(\tilde{x}^+)>M^+$ for all $j$. Define $\tilde{v}=\frac{1}{\|\tilde{x}^+\|}\tilde{x}^+$ and $v_L=-\frac{2}{p_L}(\tilde{p}\cdot \tilde{v})$. Because $u_i$ is strictly concave and increasing on $\mathbb{R}^{L-1}_{++}$, we have that $g(t)=u_i(t\tilde{v})$ is increasing and $\lim_{t\downarrow 0}g'(t)>M^+$. Thus, there exists $t>0$ such that for $w^i=(0,x^i_L)+tv$, $U_i(w^i)>U_i(x^i)$, $w^i_L>-s$, and $p\cdot w^i<p\cdot x^i$. This implies that $U_i(w^i)>U_i(x^{ik})$ and $p^k\cdot w^i<p^k\cdot x^{ik}$ for some $k$, which is a contradiction. Therefore, for every $i$, $u_i(\tilde{x}^i)>u_i(0)$. Choose any $\ell$ such that $p_{\ell}=0$. Because $\zeta_s$ satisfies (\ref{WL2}), there exists $i$ and $\ell'$ such that $p_{\ell'}>0$ and $x^i_{\ell'}>0$. Then, for $v^i=x^i+e_{\ell}-\varepsilon e_{\ell'}$, we have that $v^i_{\ell'}>0$ and $U_i(v^i)>U_i(x^i)$ if $\varepsilon>0$ is sufficiently small. Because $p\cdot v^i<p\cdot x^i$, we have that there exists $k$ such that $U_i(v^i)>U_i(x^{ik})$ and $p^k\cdot v^i<p^k\cdot x^{ik}$, which is a contradiction. This completes the proof in the case of first-type quasi-linear economies. Because almost the same arguments can be executed for second-type quasi-linear economies, we omit the proof of such cases. $\blacksquare$
\end{proof}

\begin{lem}\label{LEM12}
Suppose that $\xi:\mathbb{R}^L_{++}\to \mathbb{R}^L$ is a continuous function, and define
\[S^*=\{p\in \mathbb{R}^L_{++}|\|p\|=1\}.\]
Suppose also that $\xi$ satisfies the following five properties:
\begin{enumerate}[1)]
\item The function $\xi$ satisfies (\ref{WL2}) and (\ref{HDZ2}).

\item There exists $s>0$ such that $\xi_j(p)>-s$ for every $p\in \mathbb{R}^L_{++}$ and $j\in \{1,...,L\}$.

\item If $(p^k)$ is a sequence in $\mathbb{R}^L_{++}$ such that $p^k\to p\neq 0$ as $k\to 0$ and the set $J=\{j|p_j=0\}$ is nonempty, then $\|\xi(p^k)\|\to \infty$ as $k\to \infty$.

\item If $\xi(p)=0$, then $\xi$ is continuously differentiable around $p$.

\item If $\xi(p)=0$, then
\[\chi(p)=\begin{vmatrix}
D\xi(p) & p\\
p^T & 0
\end{vmatrix}\neq 0.\]
\end{enumerate}
Define $E=\xi^{-1}(0)\cap S^*$, and for $p\in E$,
\[\mbox{index}(p)=\begin{cases}
+1 & \mbox{if }(-1)^L\chi(p)>0,\\
-1 & \mbox{if }(-1)^L\chi(p)<0.
\end{cases}\]
Then, the set $E$ is finite, and
\[\sum_{p\in E}\mbox{index}(p)=+1.\]
\end{lem}

\begin{proof}
Omitted.\footnote{The original statement of this lemma is found in Proposition 5.6.1 of Mas-Colell (1985). For a modern proof, see Hosoya (2023).} $\blacksquare$
\end{proof}

\begin{lem}\label{LEM13}
Suppose that $E$ is a second-type quasi-linear economy, and let $p^*$ be an equilibrium price of this economy. Then, $f^i_L(p^*,p^*\cdot \omega^i)>0$ for every $i\in N$.
\end{lem}

\begin{proof}
First, note that if $x^i=f^i(p^*,m^i(p^*))$ and $y^j=y^j(p^*)$, then
\[\sum_{i\in N}x^i=\sum_{i\in N}\omega^i+\sum_{j\in M}y^j.\]
Therefore, we have that $(x,y)\in B_{\omega}$, and thus
\[u_i(\tilde{x}^i)\le u_i(\tilde{\omega}^i)+\alpha_i<U_i(\omega^i).\]
Because $U_i(x^i)\ge U_i(\omega^i)$, we have that $x^i_L>0$. This completes the proof. $\blacksquare$
\end{proof}

\begin{lem}\label{LEM14}
Choose a mollifier
\[\varphi(p)=\begin{cases}
Ce^{-\frac{1}{1-\|p\|^2}} & \mbox{if }\|p\|<1,\\
0 & \mbox{if }\|p\|\ge 1,
\end{cases}\]
where $C>0$ is chosen as
\[\int_{\mathbb{R}^L}\varphi(p)dp=1,\]
and for $\delta>0$, define
\[\varphi_{\delta}(p)=\delta^{-L}\varphi(p/\delta).\]
Let $\pi:\mathbb{R}^L_{++}\to \mathbb{R}$ be a $C^1$ function, and define $y(p)=\nabla \pi(p)$ and
\[\pi_{\delta}(p)=\int_{\mathbb{R}^L}\pi(p-q)\varphi_{\delta}(q)dq.\]
Then, the following results hold.
\begin{enumerate}[1)]
\item The function $\pi_{\delta}(p)$ is a $C^{\infty}$ function defined on $\{p\in \mathbb{R}^L_{++}|p_k>\delta\mbox{ for all }k\in \{1,...,L\}\}$, and for any compact set $C\subset \mathbb{R}^L_{++}$, $\pi_{\delta}$ uniformly converges to $\pi$ on $C$ as $\delta\downarrow 0$.

\item If $\pi$ is convex, then $\pi_{\delta}$ is also convex for each $\delta>0$.

\item Suppose that $\pi$ is homogeneous of degree one, and define $h_{\delta}(p)=p\cdot \nabla \pi_{\delta}(p)$. Then, there exists a constant $C'>0$ such that, if $p_k>\delta$ for all $k\in \{1,...,L\}$, then
\[\left|\frac{\partial h_{\delta}}{\partial p_k}(p)-\frac{\partial \pi_{\delta}}{\partial p_k}(p)\right|\le 2C'\max\{\|y(p)-y(p-q)\||\|q\|\le \delta\}.\]
\end{enumerate}
\end{lem}

\begin{proof}
1) is proved in many textbooks. See, for example, Theorem 7 of Appendix C in Evans (2010). 2) can easily be shown. Therefore, the remaining claim of this lemma is 3). First, recall that $e_k$ is the $k$-th unit vector. Thus,
\begin{align*}
&~h_{\delta}(p+te_k)-h_{\delta}(p)\\
=&~(p+te_k)\cdot \nabla \pi_{\delta}(p+te_k)-p\cdot \nabla \pi_{\delta}(p)\\
=&~\int_{\mathbb{R}^L}[(p+te_k)\cdot y(p+te_k-q)-p\cdot y(p-q)]\varphi_{\delta}(q)dq\\
=&~\int_{\mathbb{R}^L}[(p+te_k-q)\cdot y(p+te_k-q)-(p-q)\cdot y(p-q)]\varphi_{\delta}(q)dq\\
+&~\int_{\mathbb{R}^L}[q\cdot (y(p+te_k-q)-y(p-q))]\varphi_{\delta}(q)dq\\
=&~\int_{\mathbb{R}^L}\pi(p+te_k-q)\varphi_{\delta}(q)dq-\int_{\mathbb{R}^L}\pi(p-q)\varphi_{\delta}(q)dq\\
+&~\int_{\mathbb{R}^L}[q\cdot (y(p+te_k-q)-y(p-q))]\varphi_{\delta}(q)dq\\
=&~\pi_{\delta}(p+te_k)-\pi_{\delta}(p)+\int_{\mathbb{R}^L}[q\cdot (y(p+te_k-q)-y(p-q))]\varphi_{\delta}(q)dq.
\end{align*}
By a simple calculation,
\begin{align*}
&~\int_{\mathbb{R}^L}[q\cdot (y(p+te_k-q)-y(p-q))]\varphi_{\delta}(q)dq\\
=&~\int_{\mathbb{R}^L}[q\cdot y(p+te_k-q)]\varphi_{\delta}(q)dq-\int_{\mathbb{R}^L}[q\cdot y(p-q)]\varphi_{\delta}(q)dq\\
=&~\int_{\mathbb{R}^L}[(q+te_k)\cdot y(p-q)]\varphi_{\delta}(q+te_k)dq-\int_{\mathbb{R}^L}[q\cdot y(p-q)]\varphi_{\delta}(q)dq\\
=&~t\int_{\mathbb{R}^L}y_k(p-q)\varphi_{\delta}(q+te_k)dq+\int_{\mathbb{R}^L}[q\cdot y(p-q)](\varphi_{\delta}(q+te_k)-\varphi_{\delta}(q))dq\\
=&~t\left[\int_{\mathbb{R}^L}y_k(p-q)\varphi_{\delta}(q+te_k)dq+\int_{\mathbb{R}^L}[q\cdot y(p-q)]\frac{\partial \varphi_{\delta}}{\partial p_k}(q)dq\right]\\
&~+\int_{\mathbb{R}^L}[q\cdot y(p-q)]\left(\varphi_{\delta}(q+te_k)-\varphi_{\delta}(q)-t\frac{\partial \varphi_{\delta}}{\partial p_k}(q)\right)dq\\
=&~o(t)+t\int_{\mathbb{R}^L}y_k(p-q)(\varphi_{\delta}(q+te_k)-\varphi_{\delta}(q))dq\\
&~+t\int_{\mathbb{R}^L}y_k(p-q)\varphi_{\delta}(q)dq+t\int_{\mathbb{R}^L}[q\cdot y(p-q)]\frac{\partial \varphi_{\delta}}{\partial p_k}(q)dq\\
=&~t\int_{\mathbb{R}^L}[q\cdot y(p-q)]\frac{\partial \varphi_{\delta}}{\partial p_k}(q)dq+t\frac{\partial \pi_{\delta}}{\partial p_k}(p)+o(t).
\end{align*}
Moreover,
\begin{align*}
&~\int_{\mathbb{R}^L}[q\cdot y(p-q)]\frac{\partial \varphi_{\delta}}{\partial p_k}(q)dq\\
=&~-\int_{\mathbb{R}^L}\pi(p-q)\frac{\partial \varphi_{\delta}}{\partial p_k}(q)dq+\int_{\mathbb{R}^L}[p\cdot y(p)]\frac{\partial \varphi_{\delta}}{\partial p_k}(q)dq\\
&~-\int_{\mathbb{R}^L}[p\cdot (y(p)-y(p-q))]\frac{\partial \varphi_{\delta}}{\partial p_k}(q)dq\\
=&~-\frac{\partial \pi_{\delta}}{\partial p_k}(p)-\int_{\mathbb{R}^L}[p\cdot (y(p)-y(p-q))]\frac{\partial \varphi_{\delta}}{\partial p_k}(q)dq.
\end{align*}
To summarize the above calculations, we obtain that
\[\frac{\partial h_{\delta}}{\partial p_k}(p)=\frac{\partial \pi_{\delta}}{\partial p_k}(p)-\int_{\mathbb{R}^L}[p\cdot (y(p)-y(p-q))]\frac{\partial \varphi_{\delta}}{\partial p_k}(q)dq.\]
Now, by definition,\footnote{Note that, by the shape of $\varphi$, $C'$ is independent of $k$.}
\[\int_{\mathbb{R}^L}\delta\left|\frac{\partial \varphi_{\delta}}{\partial p_k}(q)\right|dq=\int_{\mathbb{R}^L}\delta^{-L}\left|\frac{\partial \varphi}{\partial p_k}(q/\delta)\right|dq=\int_{\mathbb{R}^L}\left|\frac{\partial \varphi}{\partial p_k}(q)\right|dq\equiv C'.\]
Moreover, by the mean value theorem, there exists $\theta_q\in [0,1]$ such that
\[\pi(p)-\pi(p-q)=y(p-\theta_qq)\cdot q.\]
Therefore, if we define
\[C_{\delta}=\max\{\|y(p)-y(p-q)\||\|q\|\le \delta\},\]
then,
\begin{align*}
&~\left|\int_{\mathbb{R}^L}[p\cdot (y(p)-y(p-q))]\frac{\partial \varphi_{\delta}}{\partial p_k}(q)dq\right|\\
\le&~\int_{\mathbb{R}^L}[|\pi(p)-\pi(p-q)-q\cdot y(p)|+|q\cdot (y(p)-y(p-q))|]\left|\frac{\partial \varphi_{\delta}}{\partial p_k}(q)\right|dq\\
\le&~\int_{\mathbb{R}^L}|q\cdot (y(p-\theta_qq)-y(p))|\left|\frac{\partial \varphi_{\delta}}{\partial p_k}(q)\right|dq+\int_{\mathbb{R}^L}|q\cdot (y(p)-y(p-q))|\left|\frac{\partial \varphi_{\delta}}{\partial p_k}(q)\right|dq\\
\le&~2C_{\delta}\int_{\mathbb{R}^L}\|q\|\left|\frac{\partial \varphi_{\delta}}{\partial p_k}(q)\right|dq\le 2C_{\delta}C',
\end{align*}
as desired. This completes the proof. $\blacksquare$
\end{proof}

\subsection{Proof of Proposition \ref{PROP1}}
Recall the definition of $B_{\omega}$. The definition of $B_{\omega}$ is
\[B_{\omega}=\{(x,y)\in \bar{A}_{\omega}|U_i(x^i)\ge U_i(\omega^i)\mbox{ for each }i\in N\}.\]
Let $\hat{\omega}^i=(\omega_1^i,...,\omega_{L-1}^i,0)$. By Lemma \ref{LEM10}, $B_{\hat{\omega}}$ is compact. By (\ref{ALPHA}),
\[\alpha_i=\max\{u_i(\tilde{x}^i)-u_i(\tilde{\omega}^i)|(x,y)\in B_{\hat{\omega}}\}<+\infty,\]
as desired. This completes the proof. $\blacksquare$

\subsection{Proof of Proposition \ref{PROP2}}
Because of Lemmas \ref{LEM5} and \ref{LEM8}, we have that $\zeta$ is single-valued if it is defined. To show (\ref{WL2}) and (\ref{HDZ2}) is easy, and thus we omit its proof. By Lemma \ref{LEM10}, we have that $B_{\omega}$ is compact. Choose any $s>\sup\{\|(x,y)\||(x,y)\in B_{\omega}\}$, and consider economy $E_s$. By Lemma \ref{LEM11} and Proposition 17.C.1 of Mas-Colell et al. (1995), we have that there exists an equilibrium price $p^*$ of $E_s$. By Lemma \ref{LEM8}, $y^j_s$ is defined on $\mathbb{R}^L_{++}$ and single-valued. Define
\[y^j=y^j_s(p^*),\ x^i=f^i_s(p^*,p^*\cdot \omega^i+\sum_{j\in M}\theta_{ij}\pi^j_s(p^*)).\]
Because $0\in Y^j$, we have that $\pi^j_s(p^*)\ge 0$, and thus,
\[U_i(x^i)\ge U_i(\omega^i).\]
Moreover,
\[\sum_{i\in N}x^i=\sum_{i\in N}\omega^i+\sum_{j\in M}y^j.\]
This implies that $(x,y)\in B_{\omega}$, and thus $x^i_L>-s$ for all $i\in N$ and $\|y^j\|<s$. By Lemma \ref{LEM8}, the latter inequality implies that $y^j=y^j(p^*)$, and thus $\pi^j_s(p^*)=\pi^j(p^*)$. Therefore,\footnote{Recall that $m^i(p)=p\cdot \omega^i+\sum_{j\in M}\theta_{ij}\pi^j(p)$.}
\[x^i=f^i(p^*,m^i(p^*)),\]
which implies that $\zeta(p^*)=0$ and $p^*$ is an equilibrium price of economy $E$. This completes the proof. $\blacksquare$

\subsection{Proof of Theorem \ref{THM1}}
First, suppose that this theorem holds for any first-type quasi-linear economy, and $E$ is a second-type quasi-linear economy. Let $\hat{E}$ be a first-type quasi-linear economy in which all components except for $\Omega_i$ are the same as that of $E$. By definition, the supply function of the $j$-th producer in economy $\hat{E}$ is the same as $y^j(p)$. Moreover, if $p^*$ is an equilibrium of $\hat{E}$, then for the excess demand function $\hat{X}^i$ of the $i$-th consumer in the economy $\hat{E}$, by Lemmas \ref{LEM2}-\ref{LEM4} and \ref{LEM13},
\[\hat{X}^i(p^*)\in \mathbb{R}^L_{++}\]
which implies that $X^i(p)=\hat{X}^i(p)$ on some neighborhood of $p^*$. By the same arguments, if $p^*$ is an equilibrium of $E$, then $X^i(p)=\hat{X}^i(p)$ on some neighborhood of $p^*$. Therefore, by our initial assumption, we obtain this theorem holds for this economy $E$. Hence, it suffices to show that this theorem holds for any first-type quasi-linear economy, and hereafter, we assume that $E$ is first-type.

We separate the proof into several steps.

\vspace{12pt}
\noindent
{\bf Step 1}. There exists $s>0$ such that the following results hold:
\begin{enumerate}[1)]
\item The set of equilibrium prices in $E$ is the same as that of $E_s$,

\item For every equilibrium price $p^*$, there exists a neighborhood $U$ of $p^*$ such that if $p\in U$, then $\pi^j(p)=\pi^j_s(p)$ and $X^i(p)=X^i_s(p)$.
\end{enumerate}

\vspace{12pt}
\noindent
{\bf Proof of Step 1}. By Lemma \ref{LEM10}, $B_{\omega}$ is compact. Hence, we can choose $s>\sup\{\|(x,y)\||(x,y)\in B_{\omega}\}$. Suppose that $p^*$ is an equilibrium price in $E$. Define $y^j=y^j(p^*)$ and $x^i=f^i(p^*,m^i(p^*))$. Then,
\[\sum_{i\in N}x^i=\sum_{i\in N}\omega^i+\sum_{j\in N}y^j.\]
Moreover, because $0\in Y^j$ for each $j$, we have that $\pi_j(p^*)=p^*\cdot y^j(p^*)\ge 0$. Therefore, $p^*\cdot x^i\ge p^*\cdot \omega^i$, and thus
\[U_i(x^i)\ge U_i(\omega^i),\]
which implies that $(x,y)\in B_{\omega}$, and thus $x^i_L>-s$ and $\|y^j\|<s$. Therefore, we have that $y^j=y^j_s(p^*)$ and $x^i-\omega^i=X^i_s(p^*)$, and thus $p^*$ is an equilibrium price in $E_s$. By the symmetrical proof, we can show that any equilibrium price in $E_s$ is that in $E$, and 1) holds.

Next, choose any equilibrium price $p^*$. By Lemma \ref{LEM8}, $y^j_s=y^j$ on some neighborhood $U'$ of $p^*$. Hence, there exists an open neighborhood $U\subset U'$ of $p^*$ such that if $p\in U$, then $X^i_{s,L}(p)+\omega^i_L>-s$. By Lemma \ref{LEM2}, we have that $X^i_s(p)=X^i(p)$ for all $p\in U$. This completes the proof of Step 1. $\blacksquare$

\vspace{12pt}
\noindent
{\bf Step 2}. For every $p\in \mathbb{R}^L_{++}$, if $y^j(q)$ is defined on some neighborhood of $p$ for all $j\in M$, then 
\begin{equation}\label{Slutsky}
\frac{\partial X^i_{\ell}}{\partial p_k}(p)=\begin{cases}
s^i_{\ell k}(p,m^i(p)) & \mbox{if }\ell\neq L,\\
s^i_{\ell k}(p,m^i(p))-\frac{X^i_k(p)-\sum_{j\in M}\theta_{ij}y^j_k(p)}{p_L} & \mbox{if }\ell=L.
\end{cases}
\end{equation}

\vspace{12pt}
\noindent
{\bf Proof of Step 2}. First, by almost the same arguments as in the proof of Step 1, we can show that for sufficiently large $s>0$, $y^j(q)=y^j_s(q)$ on some neighborhood of $p$. By Lemma \ref{LEM8}, we have that $\nabla \pi^j(q)=y^j(q)$ on this neighborhood.

If $\ell\neq L$, then by Lemma \ref{LEM6},
\[\frac{\partial X^i_{\ell}}{\partial p_k}(p)=\frac{\partial f^i_{\ell}}{\partial p_k}(p,m^i(p))=s^i_{\ell k}(p,m^i(p)),\]
as desired. If $\ell=L$, then 
\begin{align*}
\frac{\partial X^i_L}{\partial p_k}(p)=&~\frac{\partial f^i_L}{\partial p_k}(p,m^i(p))+\frac{\omega^i_k+\sum_{j\in M}\theta_{ij}y^j_k(p)}{p_L}\\
=&~s^i_{Lk}(p,m^i(p))-\frac{f^i_k(p,m^i(p))-\omega^i_k-\sum_{j\in M}\theta_{ij}y^j_k(p)}{p_L}\\
=&~s^i_{Lk}(p,m^i(p))-\frac{X^i_k(p)-\sum_{j\in M}\theta_{ij}y^j_k(p)}{p_L},
\end{align*}
as desired. This completes the proof of Step 2. $\blacksquare$

\vspace{12pt}
Fix an adjustment coefficient $(a_1,...,a_L)\gg 0$. Choose any equilibrium price $p^*$ in this economy. Note that, there exists such an equilibrium price because of Proposition \ref{PROP2}. By Step 1, we have that $y^j$ is defined and continuous on some neighborhood of $p^*$. Recall the t\^atonnement process (\ref{TP2}):
\[\dot{p}_{\ell}(t)=a_{\ell}\zeta_{\ell}(p(t)),\ p_{\ell}(0)=p_{0\ell}.\]
Because $p^*$ is an equilibrium price in this economy, we have that $p^*$ is a steady state of (\ref{TP2}). Define
\[h(p)=\sqrt{a_1^{-1}p_1^2+...+a_L^{-1}p_L^2}.\]
We note that $h(p)$ satisfies all requirements of the norm. In particular, we have that $h(ap)=ah(p)$ for every $p$ and $a>0$, and thus $Dh(p)p=h(p)$ if $p\neq 0$. Define
\[S(b)=\{p\in \mathbb{R}^L_{++}|h(p)=h(bp^*)\}.\]
Choose a sufficiently small $\varepsilon>0$ such that if $\|v\|\le \varepsilon$ and $t\in [-1,1]$, then $p^*+tv\in \mathbb{R}^L_{++}$ and for all $j\in M$, $y^j$ is defined at $p^*+tv$. Define
\[S=\{v\in \mathbb{R}^L|\|v\|=\varepsilon,\ Dh(p^*)v=0\},\]
and
\[p(t,v)=\frac{h(p^*)}{h(p^*+tv)}(p^*+tv).\]

\vspace{12pt}
\noindent
{\bf Step 3}. Define the following function
\[g^i(t,v)=\begin{cases}
\frac{1}{t^2}(p(t,v)-p^*)\cdot (X^i(p(t,v))-X^i(p^*)) & \mbox{if }t\neq 0,\\
v^TDX^i(p^*)v & \mbox{if }t=0.
\end{cases}\]
Then, $g^i$ is continuous on $[-1,1]\times S$.

\vspace{12pt}
\noindent
{\bf Proof of Step 3}. Clearly, $g^i$ is continuous at $(t,v)$ if $t\neq 0$. Therefore, it suffices to show that $g^i$ is continuous at $(0,v)$ for all $v\in S$. 

Choose any $\varepsilon'>0$. Note that, $p(0,v)=p^*$ and $h$ is continuously differentiable without $0$. We can easily check that
\begin{align}
\frac{\partial p}{\partial t}(0,v)=&~v,\label{EVAL1}\\
\frac{\partial p}{\partial v_j}(0,v)=&~0\mbox{ for all }j\in \{1,...,L\}.\label{EVAL2}
\end{align}
Define
\[q(t,v)=\|p(t,v)-(p^*+tv)\|,\ r(t,v)=(q(t,v))^2.\]
By (\ref{EVAL1}) and (\ref{EVAL2}), we have that $r(0,v)=0$, $Dr(0,v)=0$, and $D^2r(0,v)=0$ for all $v\in S$. Hence, by the formula of Taylor approximation, for all $v\in S$, there exists an open and convex neighborhood $U_v$ of $(0,v)$ such that if $(t',v'), (t'',v'')\in U_v$, then
\[|r(t',v')-r(t'',v'')|\le (\varepsilon')^2\|(t',v')-(t'',v'')\|^2/4.\]
Because
\begin{align*}
(q(t',v')-q(t'',v''))^2\le&~|(q(t',v')-q(t'',v''))(q(t',v')+q(t'',v''))|\\
=&~|r(t',v')-r(t'',v'')|,
\end{align*}
we have that if $(t',v'),(t'',v'')\in U_v$, then
\[|q(t',v')-q(t'',v'')|\le \varepsilon'\|(t',v')-(t'',v'')\|/2.\]
We can assume without loss of generality that
\[U_v=\{(t',v')\in \mathbb{R}^{L+1}||t'|<2\delta_v,\ \|v'-v\|<\delta_v\}.\]
Define
\[W_v=\{v'\in S|\|v'-v\|<\delta_v\}.\]
Then, $(W_v)$ is an open covering of $S$, and thus, there exists a finite subcovering $(W_{v_1},...,W_{v_K})$. Let $\delta^*=\min\{\delta_{v_1},...,\delta_{v_K}\}$. Then, we have that
\[\sup_{t\in ]0,\delta^*]}\frac{q(t,v)}{t}<\varepsilon'\]
for all $v\in S$.

Fix any $v\in S$. Since $f^i(ap,am)=f^i(p,m)$ and $m^i(ap)=am^i(p)$, we have that
\[X^i(p(t,v))=X^i(p^*+tv).\]
Define
\[\hat{g}^i(t,v)=\frac{1}{t}v\cdot (X^i(p^*+tv)-X^i(p^*)).\]
Then, by the chain rule and the mean value theorem, we can easily show that there exists $\delta>0$ such that if $0<|t|<\delta$ and $v\in S$, then
\[|\hat{g}^i(t,v)-v^TDX^i(p^*)v|<\varepsilon'.\]
Therefore, if $0<|t|<\min\{\delta,\delta^*\}$, we have that
\begin{align*}
|g^i(t,v)-v^TDX^i(p^*)v|\le&~|g^i(t,v)-\hat{g}^i(t,v)|+|\hat{g}^i(t,v)-v^TDX^i(p^*)v|\\
<&~\frac{1}{t}\varepsilon'\|X^i(p(t,v))-X^i(p^*)\|+\varepsilon'\\
\le&~(K+1)\varepsilon',
\end{align*}
where $K>0$ is some constant independent of $(t,v)$.\footnote{Because both $X^i(p)$ and $p(t,v)$ are continuously differentiable, their composition $X^i\circ p$ is Lipschitz on $[-1,1]\times S$.} Let $K^*>0$ be the operator norm of $DX^i(p^*)$. If $\|v-v'\|<\varepsilon'$ and $0\le |t|<\min\{\delta,\delta^*\}$, then
\begin{align*}
|g^i(t,v')-g^i(0,v)|\le&~|g^i(t,v')-g^i(0,v')|+|g^i(0,v')-g^i(0,v)|\\
<&~(K+2K^*\varepsilon+1)\varepsilon'.
\end{align*}
Therefore, $g^i$ is continuous at $(0,v)$. This completes the proof of Step 3. $\blacksquare$

\vspace{12pt}
Note that, because $S_{f^i}(p^*,m_i^*)$ satisfies (R) and (ND) in Lemma \ref{LEM7}, by Step 2, we have that $\max_{v\in S}\sum_{i\in N}g^i(0,v)<0$, and thus there exist $\delta_1>0$ and $L^+>0$ such that if $|t|\le \delta_1$, then $\max_{v\in S}\sum_{i\in N}g^i(t,v)\le -L^+$. 

\vspace{12pt}
\noindent
{\bf Step 4}. There exists $\delta>0$ such that $\delta\le \delta_1$ and if $0<|t|\le \delta$ and $v\in S$, then for all $j\in M$,\footnote{If $\mu=0$ and the economy is pure exchange, then this step is not needed.}
\[\frac{1}{t^2}(p(t,v)-p^*)\cdot (y^j(p(t,v))-y^j(p^*))\ge -\frac{L^+}{2\mu}.\]

\vspace{12pt}
\noindent
{\bf Proof of Step 4}. By the definition of $\varepsilon$ and Lemma \ref{LEM8}, there exists an open and convex neighborhood $U$ of $p^*$ such that $\pi^j(p)$ is convex and continuously differentiable on $U$, $\nabla \pi^j(p)=y^j(p)$ for all $p\in U$, and if $(t,v)\in [-1,1]\times S$, then $p^*+tv\in U$. Therefore, if $v\in S$, then
\[v\cdot \frac{\nabla \pi^j(p^*+tv)-\nabla \pi^j(p^*)}{t}\ge 0\]
for all $t\in [-1,1]\setminus\{0\}$. Choose any $v\in S$. By Taylor's theorem,
\[p(t,v)-p^*=vt+R(t,v)t^2,\]
where for $\ell\in \{1,...,L\}$,
\[R_{\ell}(t,v)=\frac{\partial^2p_{\ell}}{\partial t^2}(\theta_{\ell}t,v)/2\]
for some $\theta_{\ell}\in [0,1]$. Moreover, because $y$ is homogeneous of degree zero, we have that $y^j(p^*+tv)=y^j(p(t,v))$. Therefore,
\begin{align*}
&~\frac{1}{t^2}(p(t,v)-p^*)\cdot (y^j(p(t,v))-y^j(p^*))\\
=&~v\cdot \frac{\nabla \pi^j(p^*+tv)-\nabla \pi^j(p^*)}{t}+R(t,v)\cdot (y^j(p(t,v))-y^j(p^*))\\
\ge&~-\|R(t,v)\|\|y^j(p(t,v))-y^j(p^*)\|.
\end{align*}
This implies that there exists $\delta>0$ such that if $0<|t|\le \delta$ and $v\in S$, then
\[\frac{1}{t^2}((p(t,v)-p^*)\cdot (y^j(p(t,v))-y^j(p^*))\ge -\frac{L^+}{2\mu}.\]
We can assume that $\delta\le \delta_1$. This completes the proof. $\blacksquare$

\vspace{12pt}
\noindent
{\bf Step 5}. There exists an open neighborhood $U$ of $p^*$ such that if $p_0\in U$, then there exists a solution $p(t)$ to (\ref{TP2}) defined on $\mathbb{R}_+$, and for all such solutions, $\lim_{t\to \infty}p(t)=bp^*$ for $b=\frac{h(p_0)}{h(p^*)}$.

\vspace{12pt}
\noindent
{\bf Proof of Step 5}. First, suppose that $p(t)$ is a solution to (\ref{TP2}). Then, 
\[\frac{d}{dt}(h(p(t)))^2=\sum_{\ell=1}^La_{\ell}^{-1}p_{\ell}(t)a_{\ell}\zeta_{\ell}(p(t))=0,\]
by Proposition \ref{PROP2}. Therefore, we have that $h(p(t))=h(p_0)$ for all $t$.

Next, let $W=\{p\in \mathbb{R}^L_{++}|h(p)=h(p^*),\ h(p-p^*)<\varepsilon'\}$, where $\varepsilon'>0$ is so small that for all $p\in W$, there exists $v\in S$ and $t\in [0,\delta]$ such that $p$ is proportional to $p^*+tv$. Let $U=\{p\in\mathbb{R}^L_{++}|(h(p^*)/h(p))p\in W\}$. Define
\[V(p)=(h(p-(h(p)/h(p^*))p^*))^2.\]
Then, by Steps 3-4, we have that when $p(t)$ is not proportional to $p^*$, then
\[\frac{d}{dt}V(p(t))<0\]
for every solution $p(t)$ to (\ref{TP2}) such that $p_0\in U$. Because $V(p)>0$ if $p$ is not proportional to $p^*$ and $V(bp^*)=0$ for $b>0$, we have that $V$ is a Lyapunov function of (\ref{TP2}) on $U\cap S(b)$. Therefore, if $p_0\in U$, then there exists a solution $p(t)$ to (\ref{TP2}) defined on $\mathbb{R}_+$, and for any such solution $p(t)$, $\lim_{t\to \infty}p(t)=bp^*$ for $b=\frac{h(p_0)}{h(p^*)}$. This completes the proof of Step 5. $\blacksquare$.

\vspace{12pt}
\noindent
{\bf Step 6}. Let $p^*$ be an equilibrium price such that $\|p^*\|=1$. Then, there exists a neighborhood $U$ of $p^*$ such that if $p\in U$ is an equilibrium price such that $\|p\|=1$, then $p=p^*$.

\vspace{12pt}
\noindent
{\bf Proof of Step 6}. Choose any equilibrium price $p^*$ with $\|p^*\|=1$, and let $U$ be as defined in Step 5. If there exists an equilibrium price $p_0\in U$ such that $\|p_0\|=1$ and $p_0\neq p^*$, then $p(t)\equiv p_0$ is a solution to (\ref{TP2}), which contradicts Step 5. This completes the proof of Step 6. $\blacksquare$

\vspace{12pt}
\noindent
{\bf Step 7}. Define $S^*=\{p\in \mathbb{R}^L_{++}|\|p\|=1\}$. Then, there uniquely exists $p^*\in S^*$ that is an equilibrium price of this economy.

\vspace{12pt}
\noindent
{\bf Proof of Step 7}. Choose $s>0$ in Step 1. Let $E^*$ be the set of all equilibrium prices in economy $E_s$ whose norm is $1$. Because of Lemma \ref{LEM11}, we have that $E^*$ is compact, and by Step 6, it is finite. Choose $\varepsilon>0$ such that for any $p^*\in E^*$, if $\|p-p^*\|\le 4\varepsilon$, then $p\in \mathbb{R}^L_{++}$, $\pi^j(p)=\pi^j_s(p)$, $X^i(p)=X^i_s(p)$, and $\zeta(p)\neq 0$. Define
\[U_c^*=\{q\in S^*|\exists p^*\in E^*,\ \|q-p^*\|\le c\}.\]
Now, as in Lemma \ref{LEM14}, choose a mollifier
\[\varphi(p)=\begin{cases}
Ce^{-\frac{1}{1-\|p\|^2}} & \mbox{if }\|p\|<1,\\
0 & \mbox{if }\|p\|\ge 1,
\end{cases}\]
where $C>0$ is chosen as
\[\int_{\mathbb{R}^L}\varphi(p)dp=1.\]
Define
\[\varphi_{\delta}(p)=\delta^{-L}\varphi(p/\delta).\]
Construct a function $\pi^j_{\delta}$ such that
\[\pi^j_{\delta}(p)=\int_{\mathbb{R}^L}\pi^j_s(p-q)\varphi_{\delta}(q)dq,\]
and define
\[t_{p^*}(p)=\min\{1,\max\{0,2-\varepsilon^{-1}\|p/\|p\|-p^*\|\}\},\]
\[y^j_{\delta}(p)=\left(1-\sum_{p^*\in E^*}t_{p^*}(p)\right)y^j_s(p)+\sum_{p^*\in E^*}t_{p^*}(p)[\nabla \pi^j_{\delta}(p/\|p\|)+y^j_s(p^*)-\nabla \pi^j_{\delta}(p^*)].\]
Note that, $y^j_{\delta}(p)$ is continuous, homogeneous of degree zero, and $y^j_{\delta}(p^*)=y^j_s(p^*)$ for all $p^*\in E^*$. Moreover, by 1) of Lemma \ref{LEM14},
\[\lim_{\delta\to 0}\max\{|\pi^j_{\delta}(p)-\pi^j_s(p)||p\in U_{3\varepsilon}^*\}=0.\]
Because $\pi^j$ and $\pi^j_{\delta}$ are convex, by Theorem 25.7 of Rockafeller (1970), we have that
\[\lim_{\delta\to 0}\max\{\|\nabla \pi^j_{\delta}(p)-y^j_s(p)\||p\in U_{2\varepsilon}^*\}=0.\]
Therefore, $y^j_{\delta}$ converges to $y^j_s$ uniformly as $\delta\to 0$. Define
\[h^j_{\delta}(p)=p\cdot \nabla \pi^j_{\delta}(p),\]
\[m^i_{\delta}(p)=p\cdot \omega^i+\sum_{j\in M}\theta_{ij}p\cdot y^j_{\delta}(p),\]
\[X^i_{\delta}(p)=f^i_s(p,m^i_{\delta}(p))-\omega^i,\]
\[X_{\delta}(p)=\sum_{i\in N}X^i_{\delta}(p),\]
\[\zeta_{\delta}(p)=X_{\delta}(p)-\sum_{j\in M}y^j_{\delta}(p).\]
Then, $\zeta_{\delta}(p^*)=0$ for all $p^*\in E^*$, and if $\delta>0$ is sufficiently small, then $p_k>\delta$ for all $p\in U_{\varepsilon}^*$ and all $k\in \{1,...,L\}$, and $\zeta_{\delta}(p)\neq 0$ for all $p\in S^*\setminus U_{\varepsilon}^*$. Note also that, $\zeta_{\delta}(p)$ satisfies (\ref{WL2}) and (\ref{HDZ2}). Now, let $p^*\in E^*$ and choose $p\in \mathbb{R}^L_{++}$ such that $\|p\|=1$ and $\|p-p^*\|<\varepsilon$. By Lemma \ref{LEM14} and almost the same calculation as in the proof of Step 2, we can show that
\[\frac{\partial X^i_{\delta,\ell}}{\partial p_k}(p)=\begin{cases}
s^i_{\ell k}(p,m^i_{\delta}(p)) & \mbox{if }\ell\neq L,\\
s^i_{\ell k}(p,m^i_{\delta}(p))-\frac{X^i_{\delta,k}(p)-\sum_{j\in M}\theta_{ij}\left[p_k(p\cdot \nabla \pi^j_{\delta}(p))+(e_k-p_kp)\cdot \nabla h^j_{\delta}(p)\right]}{p_L} & \mbox{if }\ell=L.
\end{cases}\]
This implies that, for all $v\in \mathbb{R}^L$ such that $v\cdot p=0$,
\begin{align*}
v^TDX_{\delta}(p)v=&~\sum_{i\in N}v^TS_{f^i}(p,m^i_{\delta}(p))v-\frac{v_L}{p_L}\left[X_{\delta}(p)-\sum_{j\in M}\nabla h^j_{\delta}(p)\right]\cdot v\\
=&~\sum_{i\in N}v^TS_{f^i}(p,m^i_{\delta}(p))v-\frac{v_L}{p_L}\left[\zeta_{\delta}(p)+\sum_{j\in M}(y^j_{\delta}(p)-\nabla \pi^j_{\delta}(p))\right]\cdot v\\
&~-\frac{v_L}{p_L}\left[\sum_{j\in M}(\nabla \pi^j_{\delta}(p)-\nabla h^j_{\delta}(p))\right]\cdot v\\
=&~\sum_{i\in N}v^TS_{f^i}(p,m^i_{\delta}(p))v-\frac{v_L}{p_L}\left[\zeta_{\delta}(p)+\sum_{j\in M}(y^j_s(p^*)-\nabla \pi^j_{\delta}(p^*))\right]\cdot v\\
&~+\frac{v_L}{p_L}\left[\sum_{j\in M}(\nabla \pi^j_{\delta}(p)-\nabla h^j_{\delta}(p))\right]\cdot v.
\end{align*}
Now, let $\lambda_i^*(q,w)$ be the maximum negative eigenvalue of $S_{f^i}(q,w)$, and let
\[\lambda^*=\max\left\{\sum_{i\in N}\lambda^i(q,hm^i(q))|q\in U_{\varepsilon}^*,\ 1/2\le h\le 2\right\}.\]
Then, $\lambda^*<0$, and for sufficiently small $\delta>0$, $m^i_{\delta}(p)\in [m^i(p)/2,2m^i(p)]$ for all $p\in U_{\varepsilon}^*$, and\footnote{Note that, because $y^j(p)$ is uniformly continuous on $U_{2\varepsilon}^*$,
\[\lim_{\delta\downarrow 0}\max\{\|y^j(p)-y^j(p-q)\||p\in U_{\varepsilon}^*,\ \|q\|\le \delta\}=0.\]
Recall claim 3) of Lemma \ref{LEM14}.}
\[\sum_{j\in M}\|y^j_s(p^*)-\nabla \pi^j_{\delta}(p^*)\|<|\lambda^*|\min\{p_L|p\in U_{\varepsilon}^*\}/3,\]
\[\max_{p\in U_{\varepsilon}^*}\sum_{j\in M}\|\nabla \pi^j_{\delta}(p)-\nabla h^j_{\delta}(p)\|<|\lambda^*|\min\{p_L|p\in U_{\varepsilon}^*\}/3.\]
Suppose that $p^+\in S^*$ and $\zeta_{\delta}(p^+)=0$. By construction, there exists $p^*\in E^*$ such that $\|p^+-p^*\|<\varepsilon$, and thus $p^+\in U_{\varepsilon}^*$ and $\zeta_{\delta}$ is continuously differentiable around $p^+$. By the above calculation, we have that for all $v\in \mathbb{R}^L$ such that $v\neq 0$ and $p^+\cdot v=0$,
\begin{align*}
v^TD\zeta_{\delta}(p^+)v=&~v^TDX_{\delta}(p^+)v-\sum_{j\in M}v^TD^2\pi^j_{\delta}(p^+)(I-p^+(p^+)^T)v\\
<&~\lambda^*\|v\|^2/3-\sum_{j\in M}v^TD^2\pi^j_{\delta}(p^+)v<0,
\end{align*}
because $\pi^j_{\delta}$ is convex and thus $D^2\pi^j_{\delta}(p^+)$ is positive semi-definite. By the result in Debreu (1952), we have that
\[(-1)^L\begin{vmatrix}
D\zeta_{\delta}(p^+) & p^+\\
(p^+)^T & 0
\end{vmatrix}>0.\]
Therefore, by Lemma \ref{LEM12},
\[\sum_{p^+:\|p^+\|=1,\ \zeta_{\delta}(p^+)=0}\mbox{index}(p^+)=+1.\]
This indicates that there uniquely exists $p^+$ such that $\|p^+\|=1$ and $\zeta_{\delta}(p^+)=0$. On the other hand, by construction, we have that $\zeta_{\delta}(p^*)=0$ for all $p^*\in E^*$, which implies that $E^*$ is a singleton. This completes the proof of Step 7. $\blacksquare$.

\vspace{12pt}
Steps 5 and 7 state that our claim of Theorem \ref{THM1} is correct. This completes the proof. $\blacksquare$

\subsection{Proof of Proposition \ref{PROP3}}
Let $f(p)=\zeta_1(p,1)$. Because of (\ref{WL2}) and (\ref{HDZ2}), $(p_1^*,p_2^*)$ is an equilibrium price if and only if $f(p_1^*/p_2^*)=0$. By Theorem \ref{THM1}, there uniquely exists an equilibrium price $p^*\in \mathbb{R}^2_{++}$ with $\|p^*\|=1$. Because of the local stability, there exists an open neighborhood $U$ of $p^*$ such that if $p\in U$ and $\|p\|=1$, then $p_1/p_2<p_1^*/p_2^*$ implies that $f(p_1/p_2)>0$, and $p_1/p_2>p_1^*/p_2^*$ implies that $f(p_1/p_2)<0$. By the uniqueness of $p^*$ and the intermediate value theorem, we can set $U=\mathbb{R}^2_{++}$, and thus $p^*$ is globally stable. This completes the proof. $\blacksquare$

\subsection{Proofs in Results of Subsection 3.2}
Corollary \ref{COR1} can be easily obtained by Theorem \ref{THM2}, and thus we only show Theorem \ref{THM2}. In Lemma \ref{LEM2}, we showed that if $p_L=1$, then
\[f^i(p,m)=(\tilde{x}^i(\tilde{p}),m-\tilde{p}\cdot \tilde{x}^i(\tilde{p})),\]
where $\tilde{x}^i$ is the inverse function of $\nabla u_i(\tilde{x})$. Suppose that $c:[0,1]\to \mathbb{R}^n_{++}$ is a continuously differentiable function such that $c(0)=\tilde{p}$ and $c(1)=\tilde{q}$. Define
\[v_i(\tilde{p},m)=u_i(f^i(\tilde{p},1,m)).\]
By Roy's identity, we have that
\[\frac{\partial v_i}{\partial p_{\ell}}(\tilde{p},m)=\tilde{x}^i_{\ell}(\tilde{p}).\]
Therefore, if $p_L=q_L=1$, then
\begin{align*}
u_i(\tilde{f}^i(q,m))-u_i(\tilde{f}^i(p,m))=&~u_i(\tilde{x}^i(c(1)))-u_i(\tilde{x}^i(c(0)))\\
=&~\int_0^1\frac{d}{dt}u_i(\tilde{x}^i(c(t)))dt=\int_0^1D_{\tilde{p}}v_i(c(t),m)c'(t)dt\\
=&~\int_0^1\tilde{x}^i(c(t))\cdot c'(t)dt=V_i(\tilde{p},\tilde{q}),
\end{align*}
which completes the proof. $\blacksquare$

\section*{Acknowledgements}
I am grateful for Shinichi Suda and Toru Maruyama for their helpful lectures on t\^atonnement processes, and Takuya Masuzawa for his kindful comments on quasi-linear economies. I also thank to Shinsuke Nakamura and Chaowen Yu, who advise me for constructing the proof of our main theorem.

\section*{References}

\begin{description}
\item{[1]} Arrow, K. J., Block, H. D., Hurwicz, L. 1959. On the Stability of the Competitive Equilibrium, II. Econometrica 27, 82-109.

\item{[2]} Arrow, K. J., Debreu, G. 1954. Existence of an Equilibrium for a Competitive Economy. Econometrica 22, 265-290.

\item{[3]} Balasko, Y. 1978. Connectedness of the Set of Stable Equilibria. SIAM J. Appl. Math. 35, 722-728.

\item{[4]} Debreu, G. 1952. Definite and Semi-Definite Quadratic Forms. Econometrica 20, 295-300.

\item{[5]} Debreu, G. 1959. Theory of Value: An Axiomatic Analysis of Economic Equilibrium. Yale University Press, New Haven.

\item{[6]} Debreu, G. 1970. Economies with a Finite Set of Equilibria. Econometrica 38, 387-392.

\item{[7]} Debreu, G. 1972. Smooth Preferences. Econometrica 40, 603-615.

\item{[8]} Debreu, G. 1974. Excess Demand Functions. J. Math. Econ. 1, 15-21.

\item{[9]} Evans, L. 2010. Partial Differential Equations, 2nd ed. American Mathematical Society, Providence.

\item{[10]} Gim\'enez, E. L. 2022. Offer Curves and Uniqueness of Competitive Equilibrium. J. Math. Econ. 98, 102575.

\item{[11]} Hayashi, T. 2017. General Equilibrium Foundation of Partial Equilibrium Analysis. Palgrave Macmillan, London.

\item{[12]} Hosoya, Y. 2020. Recoverability Revisited. J. Math. Econ. 90, 31-41.

\item{[13]} Hosoya, Y. 2023. A Rigorous Proof of the Index Theorem for Economists. Commun. Econ. Math. Sci. 2, 11-30.

\item{[14]} Hosoya, Y., Yu, C. 2013. Viable Solutions of Differential Equations and the Stability of T\^atonnement Processes. Keio J. Econ. 106-1, 109-133.

\item{[15]} Kihlstrom, R., Mas-Colell, A., Sonnenschein, H. 1976. The Demand Theory of the Weak Axiom of Revealed Preference. Econometrica 44, 971-978.

\item{[16]} Mas-Colell, A. 1985. The Theory of General Economic Equilibrium: A Differentiable Approach. Cambridge University Press, Cambridge.

\item{[17]} Mas-Colell, A. 1991. On the Uniqueness of Equilibrium Once Again. In: Barnett, W. A., Cornet, B., D'Aspremont, C., Gabszewicz, J., Mas-Colell, A. (Eds.) Equilibrium Theory and Applications: Proceedings of the Sixth International Symposium in Economic Theory and Econometrics. Cambridge University Press, Cambridge, pp.275-296.

\item{[18]} Mas-Colell, A., Whinston, M. D., Green, J. R. 1995. Microeconomic Theory. Oxford University Press, Oxford.

\item{[19]} Osana, H. 1992. Foundations of Partial Equilibrium Analysis: Quasi-Linear Preferences and Consumer's Surplus. Keio J. Econ. 85-1, 31-59.

\item{[20]} Rockafeller, R. T. 1970. Convex Analysis. Princeton University Press, Princeton.

\item{[21]} Samuelson, P. A. 1950. The Problem of Integrability in Utility Theory. Economica 17, 355-385.

\item{[22]} Walras, L. 1874. Elements d' Economie Politique Pure, ou Theorie de la Richesse Sociale, Lausanne.
\end{description}
\end{document}